\documentclass[twocolumn, superscriptaddress, prb,  preprintnumbers, showpacs, floatfix, amsmath,amssymb]{revtex4-1}

\usepackage{graphicx}

\usepackage{hyperref}
\hypersetup{backref,colorlinks=true,allcolors=blue}

\usepackage{float}
\usepackage{multirow}
\usepackage{tabularx}
\usepackage[table]{xcolor}

\usepackage{tikz}
\usetikzlibrary{shapes,arrows,calc}
\usepackage{makecell}
\usepackage{cancel}
\usepackage{tcolorbox}

\usepackage[caption=false]{subfig}

\def\td#1{$_\mathrm{#1}$}
\def\tu#1{$^\mathrm{#1}$}
\def\u{DFT+\textit{U} }
\def\ud{DFT+$U^d$ }
\def\udup{DFT+$U^{d,p}$ }
\def\rut{TiO\td{2}-rutile }
\def\ana{TiO\td{2}-anatase }

\def\edit#1{#1}

\usepackage{array}
\newcommand{\PreserveBackslash}[1]{\let\temp=\\#1\let\\=\temp}
\newcolumntype{C}[1]{>{\PreserveBackslash\centering}p{#1}}
\newcolumntype{R}[1]{>{\PreserveBackslash\raggedleft}p{#1}}
\newcolumntype{L}[1]{>{\PreserveBackslash\raggedright}p{#1}}

\usepackage{pdfpages}
\usepackage{etoolbox} 

\makeatletter
\patchcmd{\@outputpage@head}{\@ifx{\LS@rot\@undefined}{}{\LS@rot}}{}{}{}
\makeatother
\begin{document}

\author{Okan K. Orhan}
\author{David D. O'Regan}
\affiliation{School of Physics, AMBER, and CRANN Institute, Trinity College Dublin, the University of Dublin, Ireland}

\title{First-principles 
Hubbard \textit{U} and Hund's \textit{J} 
corrected approximate density-functional theory
predicts an accurate fundamental gap in  rutile
and anatase TiO$_2$}

\begin{abstract}
Titanium dioxide (TiO$_2$) presents a long-standing challenge  
for approximate Kohn-Sham 
density-functional theory (KS-DFT), as well as to  its 
Hubbard-corrected extension, DFT+$U$.
We find that a previously proposed extension of first-principles DFT+$U$ 
to incorporate a Hund's $J$ correction, 
termed DFT+$U$+$J$, in combination with  
parameters calculated using a recently 
proposed linear-response theory, 
predicts  fundamental band-gaps accurate to well 
within the experimental uncertainty
in rutile and anatase TiO$_2$.
Our approach builds upon established findings that Hubbard correction to both  titanium $3d$ and oxygen 
$2p$ subspaces in TiO$_2$, symbolically giving DFT+$U^{d,p}$, 
is necessary to achieve acceptable band-gaps using DFT+$U$. 
This requirement remains when the first-principles Hund's $J$ is included.
We also find that the calculated gap  depends on the correlated subspace definition 
even when using  subspace-specific first-principles $U$ and $J$ parameters.
Using the simplest reasonable correlated subspace definition and 
underlying functional, the local density approximation, 
we show that high accuracy
results from using a relatively uncomplicated form of the  
DFT+$U$+$J$ functional.
For closed-shell systems such as TiO$_2$, we describe how 
various  DFT+$U$+$J$ functionals reduce to 
DFT+$U$ with suitably modified parameters, 
so that reliable band gaps can be calculated for rutile and anatase
with no modifications to a conventional DFT+$U$ code.

\end{abstract}

\maketitle

\section{Introduction}\label{sec:s1}
Titanium dioxide (TiO\td{2}) has been  widely used for several decades in diverse  industrial applications such as 
pigmentation and  coating~\cite{SALVADOR2000301,BRAUN1992105,MAILE2005150} due to its non-toxicity, low-cost production and thermal stability.
 TiO\td{2} came under particularly intense scrutiny with the ground-breaking work of Fujishima and Honda, who demonstrated water splitting in TiO\td{2} photo-chemical cells in the ultra-violet (UV) spectral range in 1972~\cite{FUJISHIMA1972}.
Indeed, since then,  TiO\td{2}-based structures have been engineered for diverse opto-electronic applications  such as photo-catalysts, photo-voltaics, sensors, and  for  energy and environmental applications~\cite{1347-4065-44-12R-8269,doi:10.1021/cr0500535,PELAEZ2012331}. 
In nature, TiO\td{2} has three  common polymorphs: rutile, anatase, and brookite~\cite{0957-4484-19-14-145605}. 
\rut and \ana are  more common in industrial applications,
as brookite  is  less stable and difficult to   synthesize in large volumes~\cite{doi:10.1021/cm902613h}. 
The electronic structures of pristine \rut and \ana have been extensively studied experimentally~\cite{PhysRevLett.39.1490,anatasegap,PhysRevB.18.5606,PhysRevB.51.6842,PhysRevB.52.7771},
and the most reliable  data currently available shows that  
\rut and \ana have   fundamental 
(electronic, not optical) band gaps of  
$3.03$~eV~\cite{PhysRevB.18.5606,PhysRevB.51.6842} 
and $3.47$~eV~\cite{anatasegap},  respectively.

First-principles simulations can provide  valuable insights into the
electronic structures and processes at play in 
TiO\td{2}-based systems, offering clues for the engineering of
these systems for desired applications. 
This requires the accurate description of their electronic structures in the region of their band edges, naturally, and 
this must necessarily be \edit{found}  
by means of computationally feasible and scalable methods 
if disordered structures and diverse dopants are to be assessed in any
detail.
There exist numerous acceptably reliable approaches, such as quantum chemistry methods~\cite{MACKRODT1997192,PhysRevB.51.13023}, hybrid-functionals~\cite{doi:10.1021/jp0523625}, and many-body perturbation methods~\cite{PhysRevB.82.085203,0953-8984-24-19-195503,PhysRevMaterials.3.045401}, 
\edit{but} these methods are too computationally \edit{demanding}
 for 
routine application to defective and disordered systems.  

Density-functional theory (DFT)~\cite{PhysRev.136.B864}, specifically Kohn-Sham DFT  (KS-DFT)~\cite{PhysRev.140.A1133}  using (semi-)local density exchange-correlation functionals~\cite{PhysRev.140.A1133,PhysRevB.21.5469,PhysRevB.33.8822,PhysRevB.46.6671} offers a computationally feasible framework to study the electronic structures of spatially complex TiO\td{2}-based systems. 
In the present work, with that challenge in mind, we use a
linear-scaling implementation of DFT, the 
Order-N Electronic Total Energy Package (ONETEP)~\cite{doi:10.1063/5.0004445,doi:10.1063/1.1839852,PSSB:PSSB200541457,PhysRevB.84.165131}.
However, it is well-known that semi-local KS-DFT is unable to capture 
the approximate magnitude of the band-gap of TiO\td{2}, 
a common observation among transition-metal oxides (TMOs) generally~\cite{0953-8984-9-35-010,doi:10.1063/1.2819245,GANDUGLIAPIROVANO2007219}, and so it requires,
at the very least, some corrective measures for reliable use.

In this \edit{work}, we revisit the computationally efficient approach of applying
Hubbard-model inspired corrections to approximate KS-DFT, 
namely DFT+$U$~\cite{PhysRevB.57.1505,PhysRevB.43.7570,PhysRevB.44.943,PhysRevB.48.16929,PhysRevB.50.16861,PhysRevB.52.R5467,PhysRevB.58.1201,PhysRevB.71.035105,QUA:QUA24521}
which is technically a generalized Kohn-Sham method~\cite{PhysRevB.53.3764},  
in terms of its capability of accurately describing the
fundamental electronic band gap of  TiO\td{2} polymorphs.
We find that unlike-spin Hund's $J$ correction,
specifically that introduced in the
pioneering work of Ref.~\citenum{PhysRevB.84.115108},
is the key ingredient that enables the band gaps of TiO\td{2} to be 
accurately described with this method.
A corrective functional is only as good as its parameters,
and here we use the recently-proposed minimum-tracking linear-response formalism of Ref.~\citenum{linscott2018role}
for calculating them.
Encouragingly for practical use, moreover, 
we find that for closed-shell (non-spin-polarized) systems
such as pristine TiO\td{2} and other TMOs towards the edges
of the periodic table $d$-block, 
no modification to a standard DFT+$U$ code is needed to include Hund's $J$ corrections.

No differently to what has been found in 
previous works~\cite{PhysRevB.80.233102,doi:10.1063/1.2354468,PhysRevB.80.085202,
PhysRevB.82.115109} and as an inevitable consequence of 
the O $2p$ character of the valence-band edge, 
in order to achieve significantly improved results using DFT+$U$ 
we need to apply corrective potentials to oxygen $2p$ orbitals
on the same footing as to titanium $3d$ orbitals.
The addition of Hund's $J$ does not change this fact, 
and irrespective of whether $J$ is included   
we denote this two-species correction
as DFT+$U^{d,p}$, short for
DFT+$U^d$+$U^p$, following the literature.
Unlike prior works on TiO\td{2},  in which one
or both of $U^d$ and $U^p$ was found to 
require empirical tuning for good results,  
in this work we only use first-principles
calculated $U$ and $J$ parameters
(specifically, using the 
minimum-tracking linear-response 
method~\cite{linscott2018role,glenn-phdthesis}), 
for both the Ti $3d$ and O $2p$ \edit{subspaces}.

When the unlike-spin Hund's $J$ term is included 
(using a particularly simple form of DFT+$U$+$J$, 
in agreement with
the detailed analysis  of 
Ref.~\citenum{PhysRevB.84.115108})
we predict a generalised Kohn-Sham band-gap of a   
better quality \edit{than} that which hybrid 
functionals or G$_0$W$_0$ gives, for both polymorphs,  
when gauged  against reported experimental findings
(recent,  high-quality ones in the case of anatase,
where it seems to be more challenging to measure).
\edit{The ionic geometries of
both polymorphs 
are found to be very little affected  
by the force terms due to this functional form.}
We note in passing that both functional classes, DFT+$U$ and hybrids, 
are differentiable in terms of the 
density-matrix and have a non-local potential, and so their generalised Kohn-Sham gaps
include exchange-correlation derivative 
discontinuities~\cite{doi:10.1063/1.3702391} and are directly comparable to experiment.
Promisingly for future TiO$_2$ simulation, 
and as the central conclusion of this work, 
we find that the same first-principles
DFT+$U^{d,p}$+$J^{d,p}$ \edit{functional} 
predicts the experimental fundamental 
gap to within the uncertainty of the experiment, for both polymorphs.

\section{Methodology}\label{sec:s2}
Perhaps the most well-known systematic error 
exhibited by conventional 
 approximate functionals in KS-DFT is the self-interaction error (SIE)~\cite{Cohen792,doi:10.1063/1.463297,Savin_1996,PhysRevB.56.16021,doi:10.1063/1.476859}, and its many-body generalization, 
 the delocalization error~\cite{PhysRevB.23.5048,doi:10.1063/1.2403848,doi:10.1063/1.2179072,doi:10.1063/1.2176608,doi:10.1063/1.2387954,doi:10.1063/1.2566637,doi:10.1021/cr200107z}.
SIE arises due to spurious self-repulsion of electronic density in the  KS-DFT formalism and it also persists, albeit often to a lesser extent, 
within generalized Kohn-Sham schemes.
While the origins of SIE are well understood, 
it is hard to avoid it in the 
construction of closed-form approximate functionals. 
SIE leads to the well-known significant, even drastic underestimation of 
fundamental band gaps of TMOs in particular~\cite{0953-8984-9-35-010,doi:10.1063/1.2819245,GANDUGLIAPIROVANO2007219}, 
and \rut and TiO$_2$-anatase 
are no exception in this regard~\cite{1347-4065-39-8B-L847}.
Less well understood is the generalization of SIE to account 
for the  spin degree of freedom, which is not 
necessarily less relevant in 
closed-shell systems where the spin happens
to evaluate to zero.
In this section, we outline in detail our methodology
for computing and incorporating  parameters,
the Hubbard $U^{d,p}$ for density-related error and 
Hund's $J^{d,p}$ for spin-related error, to correct a very low-cost
density functional for the specific case of TiO$_2$.

\subsection{DFT+\textit{U}+\textit{J} functionals and their simplification for closed-shell systems}\label{sec:s2s1}
DFT+$U$ is routinely applied to correct for SIE, 
particularly for the spurious delocalization of electronic states 
associated with transition-metal $3d$ orbitals.
The \u total energy is given by 
\begin{align}\label{eq:eq1}
E_\mathrm{\u}&=E_{\mathrm{DFT}}+E_U, \end{align}
where the rotationally-invariant form of 
$E_U$
for a given SIE-prone subspace~\cite{PhysRevB.52.R5467,PhysRevB.56.4900,0953-8984-9-35-010}, 
particularly if we take its relatively recent
DFT+$U$+$J$ form of Ref.~\citenum{PhysRevB.84.115108}, 
is given by
\begin{align}\label{eq:eq2}
E_U[\{n^\sigma\}] =
&\frac{1}{2}\sum_{\sigma}\sum_{m,m'}
\Big\{
\underbrace{U \left[n^{\sigma}_{mm'}\delta_{m'm}-n^{\sigma}_{mm'}n^{\sigma}_{m'm}\right]}_\mathrm{I} \nonumber \\
&-\underbrace{J\left[n^{\sigma}_{mm'}\delta_{m'm}-n^{\sigma}_{mm'}n^{\sigma}_{m'm}\right]}_\mathrm{II}  \\ \nonumber
& +\underbrace{J \left[ n^{\sigma}_{mm'}n^{\bar{\sigma}}_{m'm}\right]}_\mathrm{III}
-\underbrace{2 J\left[\delta^{\sigma \sigma_\mathrm{min}} n^{\sigma}_{mm'}\delta_{m'm}\right]}_\mathrm{IV} \Big\} .
\end{align}
Here, $\sigma$ is a spin index, $\bar{\sigma}$ is the
corresponding opposite spin, 
$\sigma_\mathrm{min}$ is the index of the minority-population spin channel
for the subspace at hand, 
$n_{mm'}$ is the subspace-projected 
KS density-matrix.
\edit{The Hubbard $U$ is, in this work at least,} interpreted
as the  subspace-and-spin-averaged net Hartree-plus-exchange-correlation interaction. 
Hund's $J$ is its spin-splitting counterpart.
We will presently detail what, precisely, is meant by 
spin-averaging and spin-splitting in this context.

The choice of appropriate form of DFT+$U$(+$J$) energy functional depends on various factors such as the system under consideration, the  limitations and robustness of approaches to determine the Hubbard $U$ and Hund's $J$ parameters, and 
\edit{the} underlying approximate density functional.
For instance, it was argued in 
Ref.~\onlinecite{PhysRevB.84.115108} that term (IV), 
which we dub the `minority spin term', 
is best not \edit{included}, as it arises due to the 
double-counting correction of a type of 
two-particle density-matrix interaction that is 
unlikely to be very much 
present in the underlying density functional.
Our numerical results will support this analysis.
It was furthermore found to lead to numerical 
instabilities, and we have also noted this  effect in our own calculations.
Our tentative explanation of this instability 
is that, when the net spin of a site is weak, the
potential arising due to this term can switch over discretely from one spin channel to the other.
The simplest functional form is achieved, of course, by neglecting the explicit correction
of exchange and effectively by setting $J=0$\edit{~eV}. 
If a value for  $J$ is available, then so is the 
Dudarev functional~\cite{PhysRevB.57.1505}, 
which includes only like-spin correction terms 
(the terms (I) and (II)) via an effective parameter, 
$U_\mathrm{eff}=U-J$ resulting symbolically in DFT+$U_\mathrm{eff}$. 

Inspired by the Dudarev model, 
we note and primarily use in this work the fact that 
 the full DFT+$U$+$J$ functional 
of Eq.~\eqref{eq:eq2} may be applied to 
closed-shell systems, without approximation, using  
an unmodified DFT+$U$ code
with no $J$ implementation.
\edit{To see this clearly, we can} rearrange Eq.~\eqref{eq:eq2} and introduce an additional parameter, $\alpha$,
which is exactly that $\alpha$ which 
is available and used to calculate the 
Hubbard $U$ in many standard DFT+$U$ codes~\cite{PhysRevB.71.035105}.
Here, it captures the inclusion minority spin term (term IV), 
when re-writing  Eq.~\eqref{eq:eq2} as
\begin{align}\label{eq:eq2.0}
E_U=\sum_{\sigma,m,m'}
\Big\{&\frac{U_\mathrm{full}}{2} \left[n^{\sigma}_{mm'}\delta_{m'm}-n^{\sigma}_{mm'}n^{\sigma}_{m'm}\right]\nonumber \\
& + \alpha \, n^{\sigma}_{mm'}\delta_{m'm}\Big\},
\end{align}
where $U_\mathrm{full}=U-2J$.
Three \edit{reasonable options} for $\alpha$ are  tested in this study, 
representing different 
\edit{interpretations} of the minority spin (term IV):

\begin{enumerate}

\item The most natural treatment (of term IV) for closed-shell systems, that suggested in Ref.~\onlinecite{PhysRevB.84.115108},  is to interpret $\sigma_\mathrm{min} = \sigma$, such that $\delta^{\sigma \sigma_\mathrm{min}}=1$.
This requires us to set $\alpha=-J/2$.

\item A modification of the latter, intended to avoid a discontinuity in the total energy at the onset of non-zero spin polarization
(it doesn't avoid such a  discontinuity in the potential),
is to ``share'' the minority spin term between the two spins, setting $\delta^{\sigma \sigma_\mathrm{min}}=1/2$ 
for closed-shell systems. This leads to  $\alpha=0$  and the resulting Hubbard functional is simply 
a Dudarev functional with $U_\mathrm{full}=U-2J$.

\item In the last case, the minority spin term is neglected, 
as it was argued \edit{that it is} best to do in its originating  
Ref.~\onlinecite{PhysRevB.84.115108}, by setting $\delta^{\sigma \sigma_\mathrm{min}}=0$.
For closed-shell systems, DFT+$U$+$J$ is then 
recovered by DFT+$U$ code with parameters $U_\mathrm{full}$
and $\alpha=J/2$.

\end{enumerate}

\edit{In this work, 
we test these different corrective functionals by 
application to both the Ti $3d$ and O $2p$ subspaces
of TiO$_2$, presenting 
DFT+$U^d$ (no O $2p$ correction) results only for the sake of illustration and completeness.
It has previously been comprehensively demonstrated,
in Ref.~\citenum{SAMAT2016891}, that it is not possible
to reconcile a reasonable band-gap with reasonable lattice
constants when applying DFT+$U$ only to Ti $3d$ subspaces
in TiO$_2$.
We further motivate our  inclusion of 
O $2p$ corrections 
in Appendix~\ref{sec:s2s2} and with
reference to Fig~\ref{FigO}.
In the Supplementary Material that accompanies
this work~\footnote{See Supplemental Material at URL,
with additional Refs.~\citenum{meagher1979polyhedral,M.1972,0953-8984-21-39-395502,0953-8984-29-46-465901,Murnaghan1944,Burdett1987,Isaak1998,Zhang2013,SWAMY2001673}, 
for detailed computational
parameters and the results of geometry optimization
using DFT+$U$+$J$.}, %
we  illustrate that the favoured  
DFT+$U$+$J$ functional 
(minority spin term neglected)
has only a very small effect on the 
lattice constants and internal ionic geometries of
both polymorphs predicted by the underlying functional.
There, we also specify the computational parameters of our
study in detail.}

\subsection{The minimum-tracking linear-response approach for first-principles Hubbard \textit{U} and  Hund's \textit{J} parameters}\label{sec:s2s3}

The results of \udup are only as good as its input Hubbard $U$ and Hund's $J$ parameters. 
Finite-difference linear-response theory provides a 
practical, widely available first-principles method for  
calculating
these~\cite{PhysRevB.58.1201,PhysRevB.71.035105,PhysRevB.84.115108}.
It has been found that linear-response tends to give Hubbard $U$ parameters for 
closed-shell systems that are too high for practical use,
and this is usually deemed to be an erroneous overestimation~\cite{PhysRevB.73.245205,doi:10.1063/1.3489110,doi:10.1063/1.4869718,linscott2018role}.
The present work provides hints that these values may be
correct after all, but that Hund's $J$ effectively reduces them and so the latter 
is (counter-intuitively, perhaps) more
important to include in closed-shell systems.
If a system has zero spin polarization, the systematic error in the approximate functional related
to the spin \emph{degree of freedom} may still be large.
In this work, we employed the recently-introduced minimum-tracking variant~\cite{glenn-phdthesis} of linear-response as implemented in the 
ONETEP \edit{DFT+$U$ implementation}~\cite{PhysRevB.85.085107,doi:10.1063/5.0004445}, 
and in particular, 
its spin-specific extension introduced in Ref.~\citenum{linscott2018role}. 
The `scaled $2\times 2$' method  was used here to 
evaluate the Hubbard $U$, Hund's $J$, and effective Hubbard 
$U$ parameters ($U_\mathrm{eff}=U-J$ and $U_\textrm{full}=U-2J$) for the 
Ti $3d$ and O $2p$ subshells of pristine \rut and \ana using %
\begin{align}
U &{}=\frac{1}{2}\frac{\lambda_U \left(f^{\uparrow\uparrow}+f^{\uparrow\downarrow}\right)+f^{\downarrow\uparrow}+f^{\downarrow\downarrow}}{\lambda_U+1} \label{eq:eq3}  \\
\mbox{and} \quad J &{}=-\frac{1}{2}\frac{\lambda_J \left(f^{\uparrow\uparrow}-f^{\downarrow\uparrow}\right)+f^{\uparrow\downarrow}-f^{\downarrow\downarrow}}{\lambda_J-1},  \label{eq:eq4}  
\end{align}
where 
\begin{align} \label{eq:eq5}  
\lambda_U=\frac{\chi^{\uparrow\uparrow}+\chi^{\uparrow\downarrow}}{\chi^{\downarrow\uparrow}+\chi^{\downarrow\downarrow}},  
\;\;\; \mathrm{and}\;\;\;
\lambda_J=\frac{\chi^{\uparrow\uparrow}-\chi^{\uparrow\downarrow}}{\chi^{\downarrow\uparrow}-\chi^{\downarrow\downarrow}},
\end{align}
and where the projected interacting response matrices are given by
$\chi^{\sigma \sigma'} = d n^\sigma / d v^{\sigma'}_\mathrm{ext}$.
The spin-dependent interaction strengths $f^{\sigma \sigma'}$
are calculated by solving $2 \times 2$ matrix equation given by
\begin{align}
f=\left[\left(\frac{\delta v_\mathrm{KS}}{\delta v_\mathrm{ext}}-1\right)\left( \frac{\delta n}{\delta v_\mathrm{ext}}\right)^{-1}\right],
\end{align}
for which matrix entities are obtained by linear fitting to small changes of  the subspace occupancies 
$\delta n^\sigma$ and subspace-averaged Kohn-Sham  potentials $\delta v_{\mathrm{KS}}^{\sigma}$ with respect to incrementally varying uniform perturbing potentials $\delta v^{\sigma}_\mathrm{ext}$ on the
targeted subspaces.
These definitions are
equivalent to a particular choice of perturbation in the  more physically transparent but perturbation-independent expressions
\begin{align} \label{tracking}  
U =\frac{ d ( v_\textrm{Hxc}^\uparrow +  v_\textrm{Hxc}^\downarrow )} {2 d (  n^\uparrow +  n^\downarrow)}  \quad
\mbox{and} \quad J =-\frac{ d ( v_\textrm{Hxc}^\uparrow -  v_\textrm{Hxc}^\downarrow )} {2 d (  n^\uparrow -  n^\downarrow)} ,
\end{align}
where the factor $1/2$ signifies averaging (or halving the of splitting  
between) the subspace averaged Hartree-plus-exchange-correlation
potentials, $v_\textrm{Hxc}^\sigma$.
Eqs.~\ref{tracking}  can be taken as 
definition of minimum-tracking linear response, and if using them
separately it is natural to use
$ \delta v^{\uparrow}_\mathrm{ext} = \delta \alpha = \delta v^{\downarrow}_\mathrm{ext}  $ for $U$ and
$ \delta v^{\uparrow}_\mathrm{ext} = \delta \beta = - \delta v^{\downarrow}_\mathrm{ext}  $ for $J$.

The scaling factors become $\lambda_U=1$ and $\lambda_J=-1$ for spin-unpolarized systems such as the pristine \rut and TiO\td{2}-anatase. 
This reflects the vanishing linear coupling between
subspace occupancy and magnetization in such systems.
As a result,  the `scaled $2\times 2$' method reduces to the  `simple $2\times 2$' method~\cite{linscott2018role},  which can be summarized as  $U = \left( f^{\sigma \bar{\sigma}} + f^{\sigma \sigma}\right) / 2 $, 
$J = \left( f^{\sigma \bar{\sigma}} - f^{\sigma \sigma}\right) / 2 $
(this gives a Dudarev $U_\textrm{eff} = f^{\sigma \sigma}$, which is reasonable  
for a like-spin-only corrective functional). 
In fact, time-reversal
symmetry can be readily exploited for closed-shell systems, where it is
sufficient to perturb one spin channel only, filling in half of the matrix elements
by symmetry, e.g. $\chi^{\uparrow\uparrow}=\chi^{\downarrow\downarrow}$.
This feature of the $2 \times 2 $ approach enabled the 
\emph{simultaneous} calculation of $U$ and $J$ in this work, 
from a single group of self-consistent calculations perturbing one
spin channel only by finite-differences.
\edit{We have verified numerically that Eqs.~\ref{tracking}
provide the same results under these conditions.}
The response \edit{matrix elements} 
coupling  Ti $3d$ and O $2p$
subspaces \edit{are not projected out}, 
as to include such entries in the response 
matrices would necessitate corresponding terms 
in the corrective functional
\edit{(these are usually called +$V$)}, which 
would complicate our analysis focused on Hund's $J$.

\section{Results and Discussion}\label{sec:s3}

\begin{table}[t!]
\renewcommand{\arraystretch}{1.5} \setlength{\tabcolsep}{12pt}
\begin{center}
{\scriptsize
\begin{tabular}{lcccc}
\hline \hline
 LDA rutile & \multicolumn{2}{c}{Ti\tu{0} conf.} & \multicolumn{2}{c}{Ti\tu{3+} conf.}\\\cline{2-5}
   &   Ti & O & Ti & O\\ \hline

 $U$  &  3.56   &   8.57  & 5.59  & 8.57\\

 $J$  &  0.29    &    0.92  & 0.38 & 0.89 \\

 $U_\mathrm{eff} = U - J$  &  3.27   &  7.66   & 5.20 & 7.68 \\

 $U_\textrm{full} = U - 2 J $  &   2.98  &   6.74 & 4.82 & 6.80 \\

 \hline \hline
\end{tabular}}
\end{center}
\caption{First-principles LDA-appropriate Hubbard $U$ and Hund's $J$ parameters 
calculated using the minimum-tracking linear-response method~\cite{glenn-phdthesis,PhysRevB.85.085107},
both for the Ti $3d$ and O $2p$ subspaces of TiO\td{2}-rutile.
The Ti $3d$ parameters depend significantly on the 
pseudo-atomic solver charge configuration used to construct the corresponding DFT+$U$
subspace, with $3+$ providing a significantly more localised subspace 
and consequently higher parameters.
Shown also are the effective Hubbard $U$ parameter of the Dudarev
model ($U_\textrm{eff}$) and that helps to reproduce   
DFT+$U$+$J$  for closed-shell systems
($U_\textrm{full}$).}
\label{tab:rut-hub}
\end{table}

We first present the calculated Hubbard $U$ and Hund's $J$ parameters
for pristine, closed-shell \rut and TiO\td{2}-anatase.
As a preliminary test, LDA-appropriate 
parameters were calculated for TiO\td{2}-rutile with
two different definitions of the DFT+$U$ target subspace for Ti $3d$ orbitals.
Specifically, both neutral and $3+$ (still non-spin-polarized)
atomic DFT calculations 
were separately performed \edit{using the
functionality available in ONETEP and described
in Ref.~\citenum{doi:10.1063/1.4728026},}
to generate pseudo-atomic orbitals to define 
the $3d$ subspace, and also
to build the initial density and NGWF guesses. 
\edit{The tensorial representation~\cite{PhysRevB.83.245124}
was used to correctly account for the  
slight nonorthogonality among the orbitals 
for a given subspace, which arises due to their
sampling in the ONETEP plane-wave-like basis.}
An OPIUM~\cite{opium} norm-conserving pseudo-potential with a $3+$ reference state was 
used  for Ti, while a charge-neutral atomic configuration was used
for O (OPIUM pseudo-potential generation, DFT+$U$ definition,
and initial density and NGWF guess generation) throughout. 
The resulting Hubbard $U$ and Hund's $J$ parameters  are summarized in Table~\ref{tab:rut-hub}.

We find that the calculated LDA Hubbard $U$ value for Ti $3d$ 
increases by $  \sim 2 $~eV or $\sim60 \%$
when going from a neutral subspace configuration to a $3+$ charge one,
due to the pronounced increase in the spatial localization of the \edit{subspace, plotted in Fig.~\ref{FigO} of Appendix~\ref{sec:s3s1s1}.}
The relatively small calculated $J$ value also increases somewhat, by
a smaller amount in multiplicative terms, $30\%$.
 $U_\textrm{full} = U - 2 J $ therefore also increases by $\sim 60 \%$.
We choose the smoother orbitals from the 
neutral pseudo-atomic solver configuration to define DFT+$U$ 
in our further calculations, \edit{and the reasoning for this will be 
discussed and demonstrated in Appendix~\ref{sec:s3s1s1}.}
There, we will see that, not only does calculating $U$ and $J$ from 
first-principles not compensate for the arbitrariness of the DFT+$U$
projectors in TiO\td{2}-rutile, it in fact \emph{reinforces} it.
We note a small but nonetheless
irksome deviation in the O $2p$ $J$ 
parameter when moving to a $3+$ Ti $3d$ NGWF
initial guess, which results from poorer
convergence characteristics when those functions are
initialised with excessive localization.

Turning next to the LDA-appropriate Hubbard $U$ and Hund's  $J$ parameters 
calculated for TiO\td{2}-anatase using the same method with a neutral Ti $3d$ subspace
definition, shown in Table~\ref{tab:ana-hub},
we note a remarkable degree of similarity with the \rut values.
In fact, the differences are within the noise of the linear-response method, 
and this reflects the similar LDA charge states 
(to well within $1\%$ for both the Ti $3d$ and  O $2p$ DFT+$U$
subspaces) and coordination chemistry in the two structures.

\begin{table}[t!]
\renewcommand{\arraystretch}{1.5} \setlength{\tabcolsep}{20pt}
\begin{center}
{\scriptsize
\begin{tabular}{lcc}
\hline  \hline  
LDA anatase  &   Ti  & O \\ \hline

 $U$  &  3.57   &    8.56  \\

 $J$  &  0.29   &    0.91 \\

 $U_\mathrm{eff} = U - J $  &  3.28 &  7.66   \\

 $U_\textrm{full} = U - 2 J $  &   3.00 &   6.75  \\

 \hline \hline  
\end{tabular}}
\end{center}
\caption{First-principles LDA-appropriate Hubbard $U$ and Hund's $J$ parameters 
calculated using the minimum-tracking linear-response method~\cite{glenn-phdthesis,PhysRevB.85.085107}, 
both for the Ti $3d$ and O $2p$ subspaces of TiO\td{2}-anatase.
Only the neutral pseudo-atomic solver configuration Ti\tu{0} is used here.
Shown also are the effective Hubbard $U$ parameter of the Dudarev
model ($U_\textrm{eff}$) and that which reproduces the 
DFT+$U$+$J$ functional (with minority term IV) for closed-shell systems
($U_\textrm{full}$).}
\label{tab:ana-hub}
\end{table}

\subsection{The first-principles band gap of  pristine TiO$_{\boldsymbol{2}}$-rutile}\label{sec:s3s1}

As a generalized Kohn-Sham theory with an differentiable density-matrix 
dependence, in same way that hybrid functionals
are~\cite{doi:10.1063/1.3702391}, the Kohn-Sham
gap of DFT+$U$ (or DFT+$U$+$J$) includes an 
explicit derivative discontinuity.
The relationship between the Kohn-Sham gap and the fundamental gap 
is thereby not only assured in principle, but the derivative discontinuity gives, 
in practice, the
opportunity for direct comparability to the experimental insulating gap.
Shown in Table~\ref{tab:rut-gap} is the  band gap of 
TiO$_{2}$-rutile
calculated using LDA and first-principles 
DFT+$U$,  DFT+$U_\mathrm{eff}$, DFT+$U_\textrm{full}$ with different $\alpha$ values, 
and explicit DFT+$U$+$J$ (minority spin term (IV) neglected), both when
applied only to the Ti $3d$ sub-shell and when applied also to the O $2p$
sub-shell.
Experimental, first-principles, semi-empirical hybrid, 
 $GW$ results, and several previous DFT+$U$ results 
 from the literature are also shown in Table~\ref{tab:rut-gap}, for comparison.
The experimental direct gap quoted~\cite{PhysRevB.18.5606,PhysRevB.51.6842} 
is based on absorption, photoluminescence, and
resonant-Raman scattering data, and is expected to be very reliable
due to the relatively small exciton binding and phonon coupling effects 
in rutile~\cite{anatasegap}, 
and moreover in light of its good agreement with available inverse photoemission 
data~\cite{PhysRevB.50.12064}.

\begin{figure}
\centering
\includegraphics[width=0.93\linewidth]{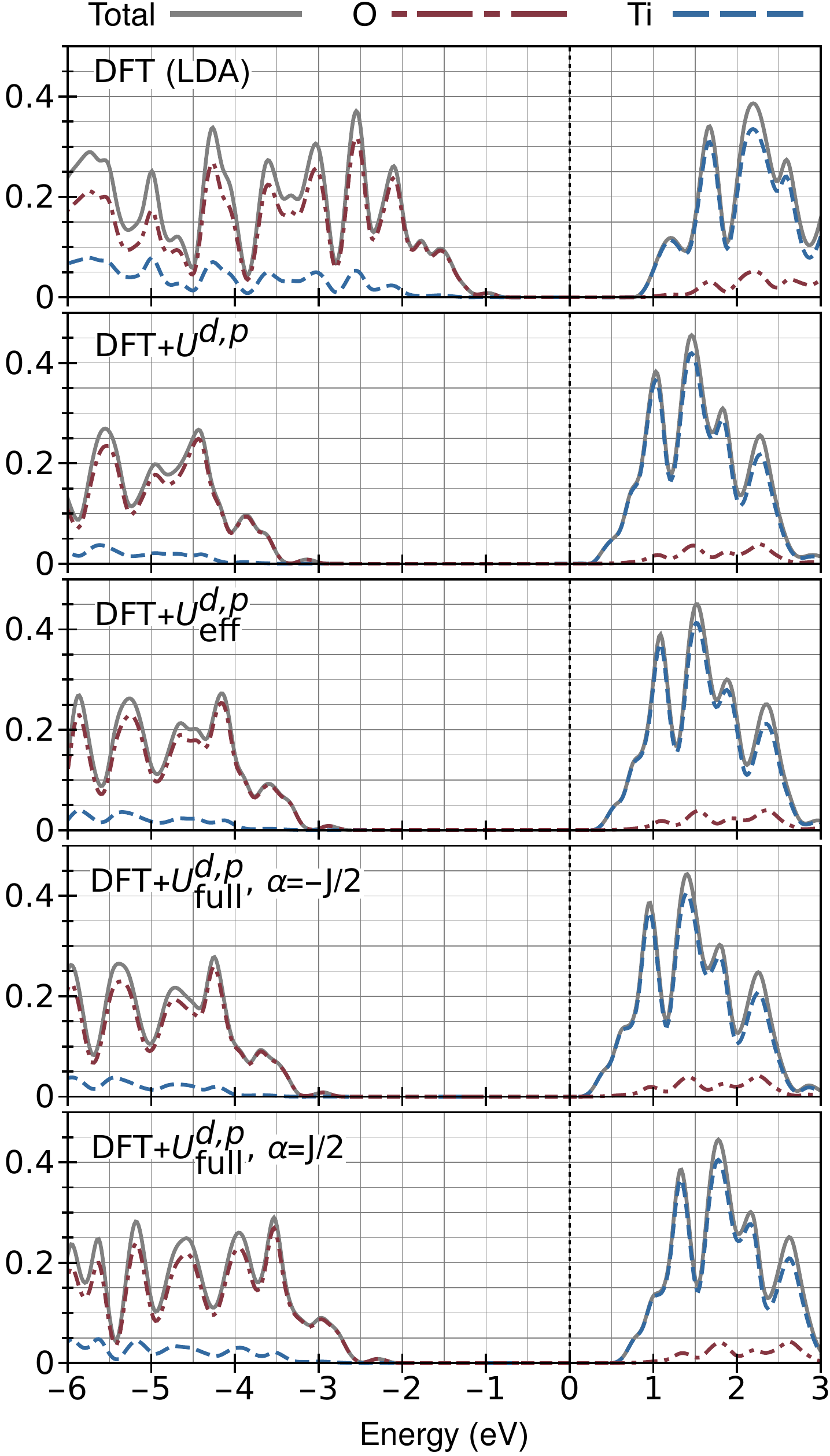}
\caption{The total and  local 
generalized Kohn-Sham density of states 
(LDOS) of pristine \rut 
calculated within  DFT(LDA), DFT+$U$ with Hund's $J$ neglected,
when treated within the Dudarev model ($U_\textrm{eff}$), and 
when treated in a matter which fully reproduces DFT+$U$+$J$
using only DFT+$U$ code for closed-shell systems ($U_\textrm{full}$), both when treated 
with ($\alpha = - J/2 $) and without ($\alpha =  J/2 $) its minority-spin (term IV). 
The spectrum is partitioned on a per-species basis using
Mulliken analysis based on the variationally optimized NGWFs.
\udup results only are  shown, using
parameters  calculated from first principles
using the minimum-tracking linear-response method, 
using only the  Ti\tu{0} pseudo-atomic solver configuration , 
and a Gaussian broadening of $0.1$~eV. 
In order to show the separate effects of the corrective
functionals tested on the valence and conduction
bands, 
each panel uses the mid-gap energy of the DFT(LDA)
calculation for $0$~eV.}
\label{Fig1}
\end{figure}

The LDA yields a Kohn-Sham band gap of $1.96$~eV, 
much lower than the experimental band gap of $3.03$~eV, 
as expected given its absence of a derivative discontinuity.
Regardless of the Hund's $J$ incorporation scheme used, 
and as is generally   attested in the literature on
calculations with $J=0$~eV, 
first-principles  DFT+$U$ applied to Ti $3d$ states only
performs poorly and here 
predicts a band gap of $2.17-2.24$~eV. 
The inadequacy of the conventional DFT+$U$ subspace definition
can be explained by comparing the 
very different valence and the conduction band edges 
characters seen in all of the local density of states
plots shown in Fig.~\ref{Fig1}, 
and  additionally motivated by
recalling the very similar degree of 
spatial localization of Ti $3d$ and O $2p$
atomic orbitals \edit{(see Fig.~\ref{FigO})}.
The valence (conduction) band edge is left almost  unaffected by applying the Hubbard correction only to the Ti $3d$ (O $2p$) sub-shell,
regardless of any reasonable Hubbard $U$ parameter
(hence, unreasonable values have been tested in the prior literature).
In qualitative agreement with that, 
we observe that the impact of the method on the band-gap increases 
substantially as soon as correction is also applied to \edit{both subshells}, within  \udup
(as we show in detail in Table~\ref{tab:rut-gap}). 

\begin{table}
\renewcommand{\arraystretch}{1.8} \setlength{\tabcolsep}{4pt}
\begin{center}
{\scriptsize
\begin{tabular}{L{1.6cm}L{1.6cm}C{1.2cm}|C{1.2cm}||C{1.2cm}|C{1.2cm}}
\hline  \hline  

\multicolumn{6}{c}{\rut $E_\mathrm{gap}$} \\ \hline

\multicolumn{2}{l}{DFT (LDA)} & \multicolumn{4}{c}{$1.96$}   \\ \cline{3-6}

  & &  \multicolumn{2}{c}{$U^d$} &  \multicolumn{2}{c}{$U^{d,p}$} \\ \cline{3-6}

\multicolumn{2}{l}{DFT+$U$}  &  \multicolumn{2}{c}{ $2.24$}  &  \multicolumn{2}{c}{ $3.59$}  \\

\multicolumn{2}{l}{DFT+$U_\mathrm{eff} = U - J $} &  \multicolumn{2}{c}{$2.21$}   &   \multicolumn{2}{c}{$3.38$} \\
 
 \multicolumn{2}{l}{DFT+$U_\textrm{full} = U - 2 J, \alpha=-J/2$} &  \multicolumn{2}{c}{$2.17$}  &    \multicolumn{2}{c}{$3.32$} \\ 
 
\multicolumn{2}{l}{DFT+$U_\textrm{full} = U - 2 J $} &  \multicolumn{2}{c}{$2.18$}  &    \multicolumn{2}{c}{$3.18$} \\ 

\multicolumn{2}{l}{DFT+$U_\textrm{full} = U - 2 J, \alpha=J/2$} &  \multicolumn{2}{c}{$2.20$}  &    \multicolumn{2}{c}{$\bf{3.04}$} \\ 

\multicolumn{2}{l}{DFT+$U$+$J$ (no minority spin term)} &  \multicolumn{2}{c}{$2.20$}  &    \multicolumn{2}{c}{$\bf{3.04}$}   \\ 

\hline

\multicolumn{4}{l}{ Experiment~\cite{PhysRevB.18.5606,PhysRevB.51.6842}} &\multicolumn{2}{c}{$\bf{3.03}$} \\

\multicolumn{4}{l}{ LDA~\cite{PhysRevB.82.115109}} &\multicolumn{2}{c}{$1.79$} \\
\multicolumn{4}{l}{ PBE~\cite{0953-8984-24-19-195503}}  & \multicolumn{2}{c}{$1.88$} \\

\multicolumn{4}{l}{ PBE~\cite{PhysRevB.86.075209}}  & \multicolumn{2}{c}{$1.86$} \\

\multicolumn{4}{l}{ PBE~\cite{PhysRevB.81.085212}}  & \multicolumn{2}{c}{$1.77$} \\

\multicolumn{4}{l}{ \edit{TB-mBJ~\cite{Gong_2012}}}  & \multicolumn{2}{c}{\edit{$2.60$}} \\

\multicolumn{4}{l}{ \edit{SCAN~\cite{doi:10.1063/1.5055623}}}  & \multicolumn{2}{c}{\edit{$2.23$}} \\

\multicolumn{4}{l}{ HSE06~\cite{PhysRevB.86.195206}} &  \multicolumn{2}{c}{$3.3$}  \\  

\multicolumn{4}{l}{ HSE06~\cite{0953-8984-24-19-195503}}& \multicolumn{2}{c}{$3.39$} \\

\multicolumn{4}{l}{ HSE06 $\left( \alpha = 0.2 \right) $~\cite{PhysRevB.81.085212}}&  \multicolumn{2}{c}{$3.05$} \\

\multicolumn{4}{l}{sX Hybrid~\cite{PhysRevB.86.075209}} & \multicolumn{2}{c}{$3.1$} \\  

\multicolumn{4}{l}{ LDA+G$_0$W$_0$~\cite{PhysRevB.82.085203}}  &  \multicolumn{2}{c}{$3.34$} \\

\multicolumn{4}{l}{ PBE+G$_0$W$_0$~\cite{0953-8984-24-19-195503}} & \multicolumn{2}{c}{$3.46$} \\
\multicolumn{4}{l}{ HSE+G$_0$W$_0$~\cite{0953-8984-24-19-195503}} & \multicolumn{2}{c}{$3.73$} \\

\multicolumn{4}{l}{ \u  ($U$=7.5 eV) \cite{0953-8984-24-20-202201}} & \multicolumn{2}{c}{$2.83$} \\
\multicolumn{4}{l}{ \u  ($U$=10 eV) \cite{doi:10.1063/1.1940739} } & \multicolumn{2}{c}{$2.97$} \\

\multicolumn{4}{l}{ \ud ($U=3.25$~eV)~\cite{doi:10.1021/jp1041316}} &   \multicolumn{2}{c}{$2.01$} \\ 

\multicolumn{4}{l}{ \udup ($U^d=3.25$~eV,  $U^p=10.65$~eV)~\cite{doi:10.1021/jp1041316}} &  \multicolumn{2}{c}{$3.67$}  \\ 

\multicolumn{4}{l}{ \udup ($U^d=3.25$~eV,  $U^p=5.0$~eV)~\cite{doi:10.1021/jp1041316}} &   \multicolumn{2}{c}{$2.69$} \\

\multicolumn{4}{l}{ \udup ($U^d=0.15$~eV,  $U^p=7.34$~eV)~\cite{PhysRevX.5.011006}} &   \multicolumn{2}{c}{$2.83$} \\

 \hline \hline  
\end{tabular}}
\end{center}
\caption{The fundamental band gap  (in eV) of  \rut 
calculated within  DFT(LDA), DFT+$U$ with Hund's $J$ neglected,
when treated within the Dudarev model ($U_\textrm{eff}$), and 
when treated in a matter which fully reproduces DFT+$U$+$J$
using only DFT+$U$ code for closed-shell systems ($U_\textrm{full}$), both when treated 
with ($\alpha = - J/2 $) and without ($\alpha =  J/2 $) its minority-spin (term IV). 
DFT+$U^d$ and \udup results are separately shown, using
parameters  calculated from first principles
using the minimum-tracking linear-response method, 
using only the neutral pseudo-atomic solver configuration Ti\tu{0}.
\edit{Prior experimental, first-principles local, semi-local, meta-generalized-gradient, 
and semi-empirical hybrid functional; perturbative G$_0$W$_0$; empirical, first-principles SCF linear-response (Ref.~\citenum{doi:10.1021/jp1041316}),} and ACBN0 (Ref.~\citenum{PhysRevX.5.011006}) DFT+$U$  values are provided for convenient comparison.
 Our central results are highlighted in bold.
 }
\label{tab:rut-gap}
\end{table}

Focusing on our own first-principles  \udup results and comparing with experiment, 
we find that when the correction for energy-magnetization curvature is neglected (letting $J=0$~eV), the band gap is overestimated by $\sim0.56$~eV with respect to the experimental  gap.
The important point here is that, even though the system harbors no magnetism in its ground-state, this does not imply that the error in the 
 approximate energy functional related to the magnetic 
 degree of freedom vanishes.
When including this effect only in the 
 like-spin term, (using Dudarev's $U_\mathrm{eff} = U - J$) this  overestimation 
 reduces to $\sim 0.35$~eV, 
 and when also applying the unlike-spin term
 (using $U_\textrm{full} = U - 2 J$ and $\alpha = - J/2 $, which is equivalent to DFT+$U$+$J$ including its standard minority spin term (IV), for  closed-shell systems such this one),
 the overestimation 
 reduces further to $\sim 0.29$~eV.

However, when we apply DFT+$U$+$J$ in its simplest
form, i.e., neglecting the minority spin term
(IV) of Eq.~\eqref{eq:eq2}
 (in practice using $U_\textrm{full} = U - 2 J$ and $\alpha = J/2 $), 
 the gap underestimation vanishes to within
the expected error in the experiment (using the zero-temperature extrapolation 
of the direct fundamental gap provided in Ref.~\onlinecite{PhysRevB.18.5606})
and the theoretical methodology.
We note that the zero-point phonon correction is held to be very small in rutile, unlike in anatase.
As shown in Table~\ref{tab:rut-gap}, we also carried out DFT+$U$+$J$ calculations using explicit
+$J$ code, with  the same results to a high precision, as predicted.
We note, in passing, that the 
deduction in the calculated gap due to the omission of the minority spin term, 
of $\sim 0.29$~eV, is very close to  
$\left( J^p - J^d \right)/2 \sim 0.31$~eV, 
as might be predicted by considering the different 
characters of the band edges and the change in the 
potentials acting upon them.

These fundamental gap changes are reflected in the local 
density of states (LDOS) plots shown in Fig.~\ref{Fig1}.
Here, we see the successive effects of first 
turning on +$U^{d,p}$ correction, and 
then by moderating it
using $J$  per Dudarev's $U_\textrm{eff} = U - J$ prescription, 
which mostly brings the
valence band back up in energy in this case.
Moving ultimately to 
DFT+$U^{d,p}_\textrm{full}$, $\alpha=J/2$ 
(which means $\alpha^d=J^d/2$, etc., and
which gives identical results to DFT+$U^{d,p}$+$J^{d,p}$
by construction), we see a further closing of the gap
and upward shift both in the valence and conduction bands.
Interestingly, we obtain an extremely similar valence-band 
DoS  from the Dudarev prescription
and 
DFT+$U^{d,p}_\textrm{full}$, $\alpha=-J/2$, 
i.e. DFT+$U^{d,p}$+$J^{d,p}$ with the minority spin term intact.
This reflects the \edit{almost-complete}
cancellation of the potentials due to
terms (III) and (IV) in Eq.~\eqref{eq:eq2}, for a
subspace near full occupancy.

\subsection{The first-principles band gap of pristine TiO$_{\boldsymbol{2}}$-anatase}\label{sec:s3s2}

\begin{table}
\renewcommand{\arraystretch}{1.8} \setlength{\tabcolsep}{0pt}
\begin{center}
{\scriptsize
\begin{tabular}{L{1.6cm}L{1.6cm}C{1.2cm}C{1.2cm}C{1.2cm}C{1.2cm}}
\hline \hline  
\multicolumn{6}{c}{TiO$_2$-anatase $E_\mathrm{gap}$} \\ \hline

\multicolumn{2}{l}{DFT (LDA)} &  \multicolumn{4}{c}{$2.21$}   \\ \cline{3-6}

&  &  \multicolumn{2}{c}{+$U^d$}  &  \multicolumn{2}{c}{+$U^{d,p}$}  \\ \cline{3-6}

\multicolumn{2}{l}{DFT+$U$} &   \multicolumn{2}{c}{$2.51$}  &  \multicolumn{2}{c}{$4.13$} \\ 

\multicolumn{2}{l}{DFT+$U_\textrm{eff} = U -  J $} &   \multicolumn{2}{c}{$2.48$}  &  \multicolumn{2}{c}{\edit{$3.88$}} \\ 
 
 \multicolumn{2}{l}{DFT+$U_\textrm{full} = U - 2 J, \alpha=-J/2$} &   \multicolumn{2}{c}{$2.41$}  &  \multicolumn{2}{c}{$3.81$} \\ 
 
\multicolumn{2}{l}{DFT+$U_\textrm{full} = U - 2 J $} &   \multicolumn{2}{c}{$2.45$}  &  \multicolumn{2}{c}{$3.65$} \\ 

\multicolumn{2}{l}{DFT+$U_\textrm{full} = U - 2 J, \alpha=J/2$} &   \multicolumn{2}{c}{$2.49$}  &  \multicolumn{2}{c}{$\bf{3.50}$} \\ 

\multicolumn{2}{l}{DFT+$U$+$J$(no minority spin term)} &   \multicolumn{2}{c}{$2.49$}  &  \multicolumn{2}{c}{$\bf{3.50}$} \\

\hline

\multicolumn{4}{l}{Experiment~\cite{anatasegap}} & \multicolumn{2}{c}{$\bf{3.47}$} \\

\multicolumn{4}{l}{PBE~\cite{0953-8984-24-19-195503}} & \multicolumn{2}{c}{$1.94$} \\

\multicolumn{4}{l}{\edit{TB-mBJ~\cite{Gong_2012}}}  & \multicolumn{2}{c}{\edit{$3.01$}} \\

\multicolumn{4}{l}{\edit{SCAN~\cite{doi:10.1063/1.5055623}}}  & \multicolumn{2}{c}{\edit{$2.56$}} \\

\multicolumn{4}{l}{HSE06~\cite{0953-8984-24-19-195503,PhysRevB.86.195206}} & \multicolumn{2}{c}{$3.60$} \\

\multicolumn{4}{l}{LDA+G$_0$W$_0$~\cite{PhysRevB.82.085203}} &  \multicolumn{2}{c}{$3.56$} \\

\multicolumn{4}{l}{PBE+G$_0$W$_0$~\cite{anatasegap}} &  \multicolumn{2}{c}{$3.61$} \\

\multicolumn{4}{l}{PBE+G$_0$W$_0$~\cite{0953-8984-24-19-195503}} &  \multicolumn{2}{c}{$3.73$} \\

\multicolumn{4}{l}{HSE+G$_0$W$_0$~\cite{0953-8984-24-19-195503}} &  \multicolumn{2}{c}{$4.05$} \\

\multicolumn{4}{l}{ \ud  ($U$=7.5 eV) \cite{0953-8984-24-20-202201}} & \multicolumn{2}{c}{$3.27$}\\

\multicolumn{4}{l}{ \ud ($U=3.23$~eV)~\cite{doi:10.1021/jp1041316}} &   \multicolumn{2}{c}{$2.43$} \\ 

\multicolumn{4}{l}{\udup ($U^d=3.23$~eV, $U^p=10.59$~eV)~\cite{doi:10.1021/jp1041316}} &  \multicolumn{2}{c}{$4.24$}  \\ 

\multicolumn{4}{l}{\udup ($U^d=3.23$~eV,  $U^p=5.0$~eV)~\cite{doi:10.1021/jp1041316}} &   \multicolumn{2}{c}{$3.23$} \\

 \hline \hline  
\end{tabular}}
\end{center}
\caption{The  band gap  (in eV) of  \ana 
calculated within  DFT(LDA), DFT+$U$ with Hund's $J$ neglected,
when treated within the Dudarev model ($U_\textrm{eff}$), and 
when treated in a matter which fully reproduces DFT+$U$+$J$
using only DFT+$U$ code for closed-shell systems ($U_\textrm{full}$), both when treated 
with ($\alpha = - J/2 $) and without ($\alpha =  J/2 $) its minority-spin (term IV).
DFT+$U^d$ and \udup results are separately shown, using
parameters  calculated from first principles
using the minimum-tracking linear-response method, 
using only the neutral pseudo-atomic solver configuration Ti\tu{0}.
\edit{Prior experimental, first-principles local, semi-local, meta-generalized-gradient, 
and semi-empirical hybrid functional; perturbative G$_0$W$_0$; empirical and}  first-principles SCF
 linear-response DFT+$U$ (Ref.~\citenum{doi:10.1021/jp1041316})
 values from the literature are provided for convenient comparison.
 Our central results are highlighted in bold.}
\label{tab:ana-gap}
\end{table}

\begin{figure}
\centering
\includegraphics[width=0.93\linewidth]{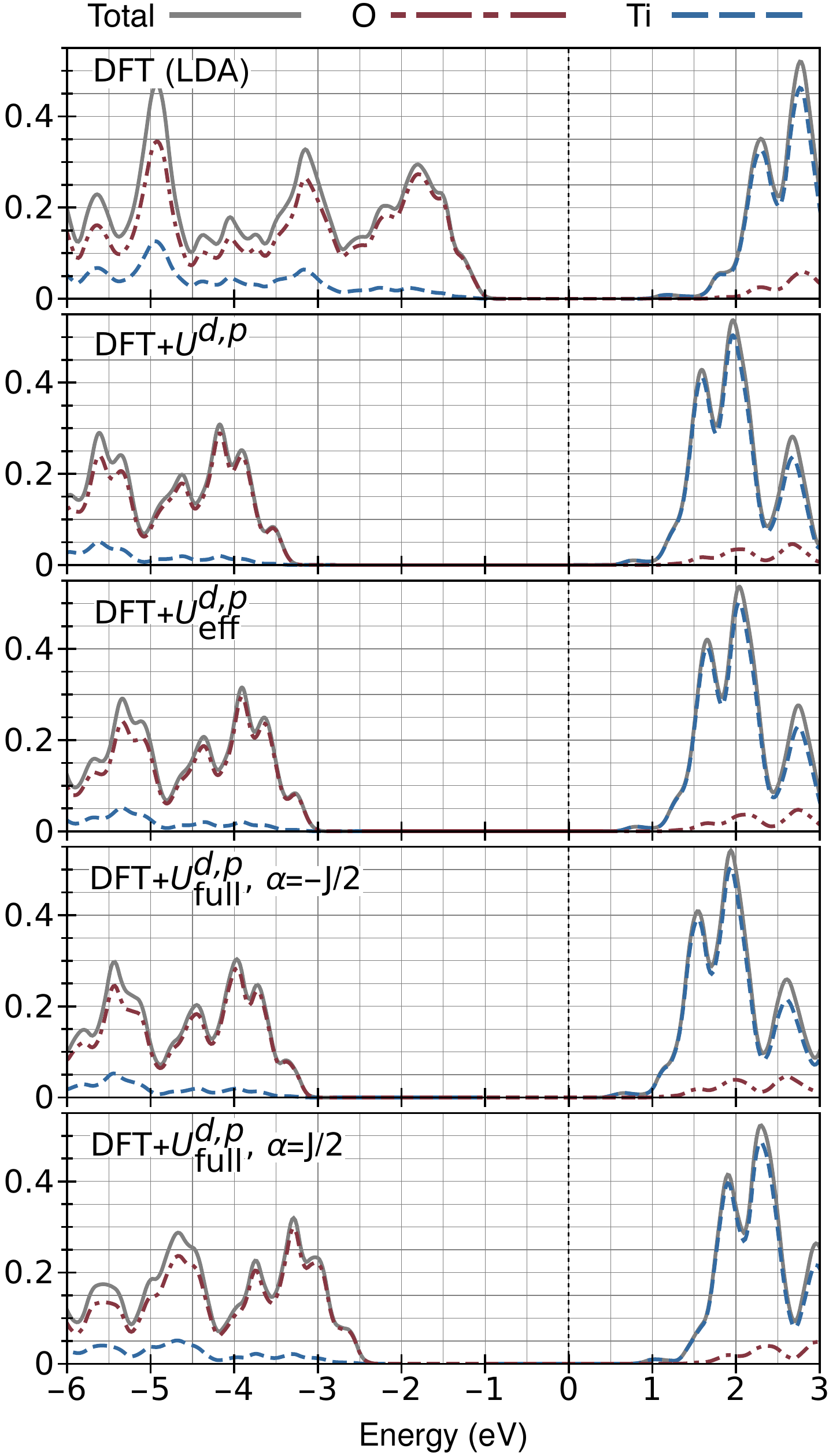}
\caption{The total and  local 
generalized Kohn-Sham density of states 
(LDOS) of pristine \ana 
calculated within  DFT(LDA), DFT+$U$ with Hund's $J$ neglected,
when treated within the Dudarev model ($U_\textrm{eff}$), and 
when treated in a matter which fully reproduces DFT+$U$+$J$
using only DFT+$U$ code for closed-shell systems ($U_\textrm{full}$), both when treated 
with ($\alpha = - J/2 $) and without ($\alpha =  J/2 $) its minority-spin (term IV). 
The spectrum is partitioned on a per-species basis using
Mulliken analysis based on the variationally optimized NGWFs.
\udup results only are  shown, using
parameters  calculated from first principles
using the minimum-tracking linear-response method, 
using only the  Ti\tu{0} pseudo-atomic solver configuration , 
and a Gaussian broadening of $0.1$~eV. 
In order to show the separate effects of the corrective
functionals tested on the valence and conduction
bands, 
each panel uses the mid-gap energy of the DFT(LDA)
calculation for $0$~eV.}
\label{Fig3}
\end{figure}

A similar procedure was followed for  pristine \ana 
as that which we have outlined for TiO$_2$-rutile, except that
only the neutral atomic configuration of Ti was used in the 
pseudo-atomic solver, in view of our previously discussed findings.
As reflected in the calculated $U$ and $J$ parameters of
Tables~\ref{tab:rut-hub} and \ref{tab:ana-hub}, the electronic structures of the two
polymorphs are rather similar, and again the valence (conduction) 
band edge is dominated by O $2p$ (Ti $3d$) character in TiO$_2$-anatase, 
necessitating \udup for successful gap correction.
Shown in Table~\ref{tab:ana-gap} is the fundamental band gap of 
TiO$_2$-anatase
calculated using LDA and first-principles 
DFT+$U$,  DFT+$U_\mathrm{eff}$, DFT+$U_\textrm{full}$, 
and  DFT+$U$+$J$ (minority spin term (IV) included, spin-averaged, and neglected), both when
applied only to the Ti $3d$ sub-shell and when applied also to the O $2p$
sub-shell.
The corresponding NGWF-partitioned Mulliken
LDOS plots are show in in Fig.~\ref{Fig3}.
We anticipate a slight overestimation 
in our calculated gap values for 
TiO$_2$-anatase, 
due to our 
necessarily finite effective 
sampling of the Brillouin zone.
The band gap of  anatase is of indirect 
character and, 
\edit{while our sampling is chosen 
to closely sample the LDA
band edges,
 we cannot be guaranteed to precisely 
sample the} valence band maximum
(most studies hold the fundamental gap of rutile to be direct at $\Gamma$, on the other hand, which we do sample).
Again, experimental, first-principles, semi-empirical hybrid, 
many-body perturbation theory, and several previous DFT+$U$ results 
from the literature are  shown for comparison.

While anatase has been thoroughly studied using optical 
techniques~\cite{doi:10.1063/1.5043144},
our focus here is on the fundamental electronic gap.
For the latter, very little direct data is available, but fortunately there has 
recently been reported 
angle-resolved photoemission spectroscopy with
n-type doping (to circumvent the need for
\edit{inverse} photoemission) in Ref.~\citenum{anatasegap},
strongly supported by temperature-dependent 
many-body perturbation theory calculations including electron-phonon coupling.
The fundamental gap reported in the latter work is higher than that found elsewhere
in older studies, and the 
reason is that, whereas the commonplace mis-identification between the optical and fundamental gap is
not very significant for rutile (the exciton binding is $\sim 4$~meV), 
 it is not at all reasonable for anatase, which \edit{is reported to 
exhibit} relatively very large exciton binding $\sim 0.18$~eV effects 
in its low-energy optical spectra~\cite{anatasegap}.

The LDA gives a Kohn-Sham band gap of $2.21$~eV, substantially underestimating 
the experimental electronic gap of $3.47$~eV.
\ud  is ineffective at opening the gap as 
is in TiO$_2$-rutule, given the 
LDA-appropriate calculated first-principles $U$ 
and $J$ parameters.
 \udup opens the  gap very efficiently and,
 closely mirroring what we found for TiO$_2$-rutile, 
 both DFT+$U$ with $J$ neglected
 and Dudarev's DFT+$U_\textrm{eff}$ cause the gap to be overestimated.
Similarly, again, first-principles DFT+$U$+$J$ including O $2p$
correction gives decent agreement with the experimental gap, overestimating it by $0.03$~eV ($0.34$~eV)
when the minority spin term is neglected (included).
Interestingly, both the HSE06 and DFT+G$_0$W$_0$
approximations seem to  better 
recover the anatase gap than the rutile
one, based on the available literature. 
DFT+$U_\textrm{full}$, $\alpha = J/2$ (which is to say, technically,  
first-principles 
DFT+$U^{d,p}$+$J^{d,p}$ with the minority spin term
neglected, which doesn't require an explicit 
Hund's $J$ implementation for closed-shell systems) seems to be
very competitive with respect to both methods as far as both the 
fundamental gap and  computational complexity are concerned.
The key ingredient for TiO$_2$ in this sort of method, aside
from the established message that the
 O $2p$ subspace needs to be treated on the same
footing as the Ti $3d$ one, is evidently to correct both for 
the usual charge-related ($U$) and
spin-related ($J$) systematic errors in the approximate functional.
Indeed, more generally it has been shown in 
Ref.~\citenum{linscott2018role}, by using the
$2 \times 2$ formalism to analyse the linear-response 
approach for Hubbard $U$ parameter calculation,
that the non-neglect of Hund's $J$ is advisable even on 
abstract consistency grounds.

\section{Conclusions}\label{sec:s4}

We have shown that the DFT+$U$+$J$ functional developed in Ref.~\citenum{PhysRevB.84.115108},
in combination with the first-principles procedure for calculating $U$ and $J$ parameters developed
in Ref.~\citenum{linscott2018role}, yields fundamental gaps that are in very close agreement with the
most sophisticated available zero-temperature-approaching experimental findings for TiO$_2$.
The residual errors, $0.01~$eV for rutile and $0.03~$eV for anatase, are within the anticipated errors due to factors
such as neglected zero-point phonon motion and relativistic effects, the pseudopotential approximation, imperfect 
Brillouin zone sampling (more relevant for anatase), and various sources of experimental uncertainty.
Interestingly, the method performs better than both hybrid functionals and perturbative G$_0$W$_0$
for the fundamental gap, while retaining a semi-local DFT-like level of computational cost
(even linear-scaling~\cite{PhysRevB.85.085107}, algorithmically, though we don't exploit that here).

An important and surprising finding of this work, 
\edit{which we go on to discuss in Appendix~\ref{sec:s3s1s1},}
is that, contrary to our expectation, the first-principles calculation of $U$ and $J$ for 
TiO$_2$-rutile  acts to reinforce the 
numerically-significant  arbitrariness~\cite{PhysRevB.82.081102}
of DFT+$U$ with respect to the (too often unstated) choice 
of localized orbitals defining the  subspaces targeted for correction.
The good news here is that it is the default, neutrally-charged, isolated atomic configuration that yields the accurate gaps.
In our experience to date, the introduction of chemical intuition when defining atomic solver charge states
for DFT+$U$ projector construction yields worsened results together with worsened convergence behaviour.

We judge that our results are, overall, very encouraging for the continued, very widespread use of DFT+$U$ and its extensions
 for
studying TiO$_2$, and that they serve as a counter-example
to the concept that such methods are  fundamentally limited in their applicability to high-spin systems.
It remains for a future study to establish whether TiO$_2$
is a special case for the 
Ref.~\citenum{PhysRevB.84.115108}
and
Ref.~\citenum{linscott2018role} combination, 
or whether it is as successful for oxides, particularly
closed-shell oxides, more generally.
What has hampered closed-shell 
applications to date, as highlighted in Ref.~\citenum{C7CP00025A}, 
 have been available Hubbard $U$ values, calculated or otherwise, that are too high for practical use.
Our results demonstrate that Hund's $J$, which is subtracted from $U$ once in the Dudarev formalism, 
and effectively \emph{twice} in DFT+$U$+$J$ for closed-shell systems, yielding
$U_\textrm{full} = U - 2 J $,  may be
 the key ingredient to moderating  the $U$.
The first-principles $U$ values in 
common circulation for Ti $3d$ orbitals in TiO$_2$, 
in the range of approximately $3-4$~eV depending
on the projector choice,  
are perhaps fine after all.
Meanwhile, our directly calculated, relatively high-seeming-at-first $U$ values for O $2p$ orbitals in TiO$_2$
(which are more localised than Ti $3d$ ones, \edit{see the plot in  Fig.~\ref{FigO})} sit  among the few previously 
reported calculated values for TiO$_2$ in the literature~\cite{PhysRevB.78.241201,PhysRevX.5.011006}.

Our results are consistent with the prescriptions detailed in Ref.~\citenum{PhysRevB.84.115108}
and Ref.~\citenum{linscott2018role}, for the use and calculation of $U$ and $J$ parameters, respectively, 
being correct.
The contribution of the explicit unlike-spin $J$ correction (term (III) in Eq.~\eqref{eq:eq2}) to the potential 
subspace matrix elements for spin $\sigma$,  is given  by
$ V^{J \sigma}_{m m'} = J  n^{\bar{\sigma}}_{m m'} $.
It seems that this is a very good approximation, 
given that there are $J$ parameters involved for two different subspace types and the net
result is very accurate \edit{as far as the gap is concerned}.
Our results strongly support the conclusions of Ref.~\citenum{PhysRevB.84.115108} 
that the minority spin term (IV) of Eq.~\eqref{eq:eq2}, which arises only due to 
the double-counting correction of a unlike-spin interaction that unlikely to be
well described in the underlying functional in the first place, should be neglected.
Equivalently, they support the conclusion that 
the fully localized limit double-counting term of Refs.~\citenum{Anisimov_1997}
and ~\citenum{PhysRevB.49.14211} 
is sufficient at this level of theory, 
at least as far as the potential is concerned.
The DFT+$U$+$J$ gap is just one aspect of the potential, 
of course, 
and its correctness cannot be used to judge
whether the double-counting in the total energy is correct, for example.
In previous works, we have pointed out cases where
the standard DFT+$U$ potential fails due to non-satisfaction of
Koopmans' condition~\cite{PhysRevB.94.220104}, or due to inadequate 
projection onto the states adjacent to the 
band edges~\cite{PhysRevB.99.165120}, 
neither of which effects are expected to be  alleviated 
particularly by
the incorporation of  Hund's $J$.

On a similar cautionary note, 
it is worth emphasising that our first-principle 
calculations of $U$ and $J$ in TiO$_2$
were made simpler  by the vanishing occupancy-magnetization coupling in closed-shell systems,
by which we mean that $ d \left( n^\uparrow + n^\downarrow \right) / d \beta = 0 = 
d \left( n^\uparrow - n^\downarrow \right) / d \alpha$.
In this case, the elegant  
formulae of of Eq.~\eqref{tracking} 
become unambiguous with respect to the spin-polarization
of the perturbing potential.
In our current view, these two formulae are essentially the correct ones for $U$ and $J$,  
neglecting self-consistency over parameters.
As a result, without approximation and very conveniently, 
we were able to perturb one 
spin only and obtain $U$ and $J$ simultaneously. 
A disadvantage of this decoupling, however, 
is that we cannot judge on the basis 
of the present calculations
between the merits of the ``scaled $2 \times 2$'' and 
``simple $2 \times 2$'' procedures
of Ref.~\citenum{linscott2018role}, since they become identical.
Overall,  there is without doubt much further work to be done on developing self-contained 
corrective techniques such as first-principles DFT+$U$+$J$ for approximate density-functional theory, 
which side-step the evolution of increasingly costly closed-form functionals.
Meanwhile, our results here may prove to significantly lower the computational barrier to simulating
accurate spectral quantities in large, possibly defect-containing or disordered 
super-cells of TiO$_2$.

\section{Acknowledgements}

\edit{We wish  to thank 
Glenn Moynihan, 
Edward Linscott, 
Gilberto Teobaldi, and
Harald Oberhofer 
for helpful discussions.}
We acknowledge the support of 
Trinity College Dublin School of Physics, 
Science Foundation Ireland (SFI)
through The Advanced Materials and Bioengineering Research Centre 
(AMBER, grant 12/RC/2278 and 12/RC/2278\_P2), and  the European Regional Development Fund (ERDF).
We also acknowledge the DJEI/DES/SFI/HEA Irish Centre for High-End Computing (ICHEC) 
for the provision of computational facilities and support.
We further acknowledge Trinity Centre for High Performance Computing
and Science Foundation Ireland, for the maintenance and
funding, respectively, of the Boyle (Cuimhne upgrade) cluster on which
further calculations were performed.

\appendix

\section{DFT+\textit{U} on 2p and 3d orbitals: DFT+$\boldsymbol{U^{d,p}}$ }\label{sec:s2s2}

\begin{figure}
\centering
\includegraphics[width=\linewidth]{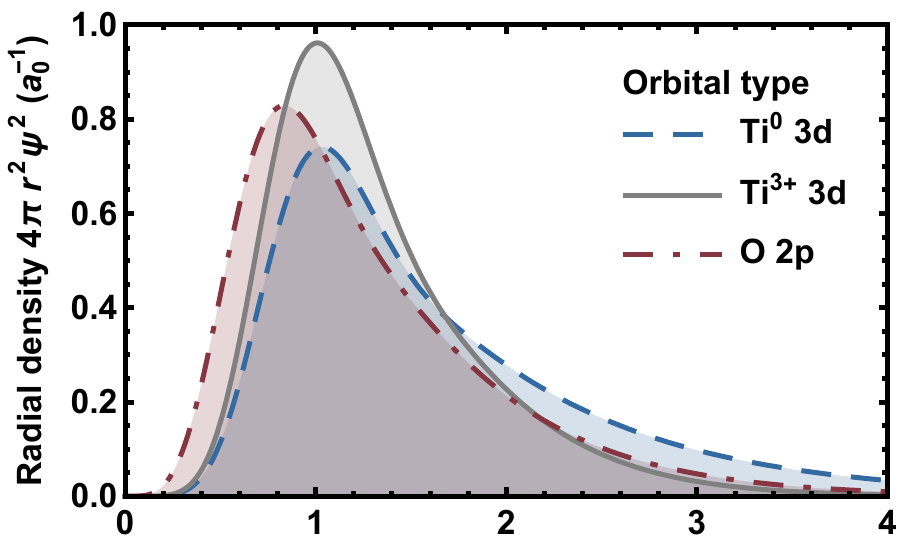}
\caption{\edit{The radial probability distributions of
the three pseudo-orbital types used to define DFT+$U$(+$J$)
subspaces in this work, as defined in the main text.
The oxygen $2p$ subspaces
are more localised than their
titanium $3d$ counterparts, and thus are  
reasonable candidates for
correction testing.
The Ti pseudo-atomic solver charge state  
significantly 
affects the localization of the $3d$ orbitals, and 
the resulting gaps.}}
\label{FigO}
\end{figure}

In principle, SIE is harboured by all subshells
and cannot be partitioned out between them, 
however, it is commonly more dominant in 
$3d$ subshells due to their 
spatially  localized nature. 
Hence, in titanium-comprising systems, the Hubbard correction in \u  is conventionally 
applied to the Ti $3d$ subshell only.
The Hubbard $U$ parameters used for the $3d$ orbitals of Ti atom have ranged over $\sim 2.5 - 10$ eV~\cite{doi:10.1021/jp111350u}, and have most
commonly been determined by tuning  to some observed quantity~\cite{doi:10.1063/1.2996362,PhysRevB.53.1161,PhysRevB.75.195212,PhysRevB.77.045118,PhysRevB.77.235424,PhysRevB.78.241201,PhysRevB.81.033202}.
Even when overlooking our serious concerns regarding \edit{the}  robustness and conceptual
validity of $U$ \edit{value} calibration to observable quantities, 
particularly when those are not ground-state observables, 
a practical problem arises  for DFT+$U$ due to 
the location of Ti on the extreme left of the transition-metal block.
It is well known that Hubbard $U$ correction to the $3d$-orbitals alone is not very effective
for opening the band gap of TiO$_2$, which saturates even with unreasonably large $U$ values, 
as the dominant $2p$-states at the valence band-edge remain barely affected.
Moreover, when \edit{actually plotted, as they are 
in Fig.~\ref{FigO}} the $2p$ pseudo-orbitals of O atoms are 
are \edit{rather} \emph{more} localized than their Ti $3d$ counterparts, 
and so it is not at all unreasonable, quite the contrary, to calculate
(or at least tune, where calculation is not possible) Hubbard $U$ and even
Hund's $J$ parameters for O $2p$.
Indeed, it has been demonstrated in several prior works that
applying the Hubbard $U$ correction simultaneously  on the 
$3d$ orbitals of Ti and the $2p$ orbitals of O atoms,  symbolically giving DFT+$U^{d,p}$,
readily addresses the aforementioned gap saturation problem and
provides a more accurate description of the band structure around the Fermi level~\cite{PhysRevB.80.233102,doi:10.1063/1.2354468,PhysRevB.80.085202,PhysRevB.82.115109}.

\section{The effects 
\edit{on the density of states} 
of the \edit{choice of} pseudo-atomic solver 
configuration for generating
the \edit{Ti 3d} DFT+$\boldsymbol{U}$ subspace}\label{sec:s3s1s1}

\begin{table}[h]
\renewcommand{\arraystretch}{1.8} \setlength{\tabcolsep}{0pt}
\begin{center}
{\scriptsize
\begin{tabular}{L{1.6cm}L{1.6cm}C{1.2cm}|C{1.2cm}||C{1.2cm}|C{1.2cm}}
\hline  \hline  

\multicolumn{6}{c}{\rut $E_\mathrm{gap} $} \\ \hline

\multicolumn{2}{l}{Subspace definition} & \multicolumn{2}{c}{Ti\tu{0}} &  \multicolumn{2}{c}{Ti\tu{3+}} \\ \cline{3-6}

\multicolumn{2}{l}{DFT(LDA)} & \multicolumn{2}{c}{$1.96$}  & \multicolumn{2}{c}{$1.96$} \\ \cline{3-6}

  & & +$U^d$ & +$U^{d,p}$&   +$U^d$  & +$U^{d,p}$ \\ \cline{3-6}

\multicolumn{2}{l}{$U$} & $2.24$  &   $3.59$ & $2.69$ & $4.20$ \\ 

\multicolumn{2}{l}{$U_\textrm{eff} = U - J$} & $2.21$  &   $3.38$ & $2.63$ & $3.94$ \\ 

\multicolumn{2}{l}{$U_\textrm{full} = U - 2 J, \alpha=-J/2$} &  $2.17$  &   $3.32$ & $2.52$ & $3.81$ \\ 

\multicolumn{2}{l}{$U_\textrm{full}= U - 2 J $ from Ti\tu{0}} & $2.18$  &   $3.18$ & $\bf{2.31}$ & $\bf{3.33}$\\ 

\multicolumn{2}{l}{$U_\textrm{full}= U - 2 J $ from Ti\tu{3+}} & $\bf{2.38}$  & $\bf{3.46}$  & $2.57$ & $3.69$ \\ 

\multicolumn{2}{l}{$U_\textrm{full} = U - 2 J, \alpha=J/2$} &  $2.20$  &  $3.04$ & $ 2.62$ & $3.58$ \\ 

\multicolumn{2}{l}{$U$+$J$ (no minority spin term)} & $2.20$  &  $3.04$ & $2.64$ & $3.58$    \\

 \hline \hline  
\end{tabular}}
\end{center}
\caption{This table highlights the arbitrariness of DFT+$U$ with respect to the
targeted subspace choice, which is not compensated for in this system 
by first-principles calculation of the $U$ and $J$ parameters.
Shown is the  band gap (in eV) of  \rut 
calculated within  DFT(LDA), DFT+$U$ with Hund's $J$ neglected,
when treated within the Dudarev model ($U_\textrm{eff}$), and
when treated in a matter which fully reproduces DFT+$U$+$J$
using only DFT+$U$ code ($U_\textrm{full}$).
DFT+$U^d$ and \udup results are separately shown, and these depend
on the pseudo-atomic solver configuration (neutral or $3+$) used
to define the targeted Ti $3d$ subspace, together with the corresponding 
subspace-dependent $U$ and $J$ parameters.
For the intermediate case of $U_\textrm{full}$ with $\alpha = 0$, i.e., DFT+$U$+$J$ with its minority term split over the two spins,
we show the effect on the gap of separately changing the subspace used to calculate
the parameters, and the subspace used to apply the parameters, revealing that these effects
combine to reinforce, not to cancel, the subspace-dependence in this system.
The gaps from ``mismatched'' calculations, with parameters from the other subspace type, 
are shown in  bold.
}
\label{tab:rut-gap-Ti-conf-anal}
\end{table}

\begin{figure}
\centering
\includegraphics[width=0.93\linewidth]{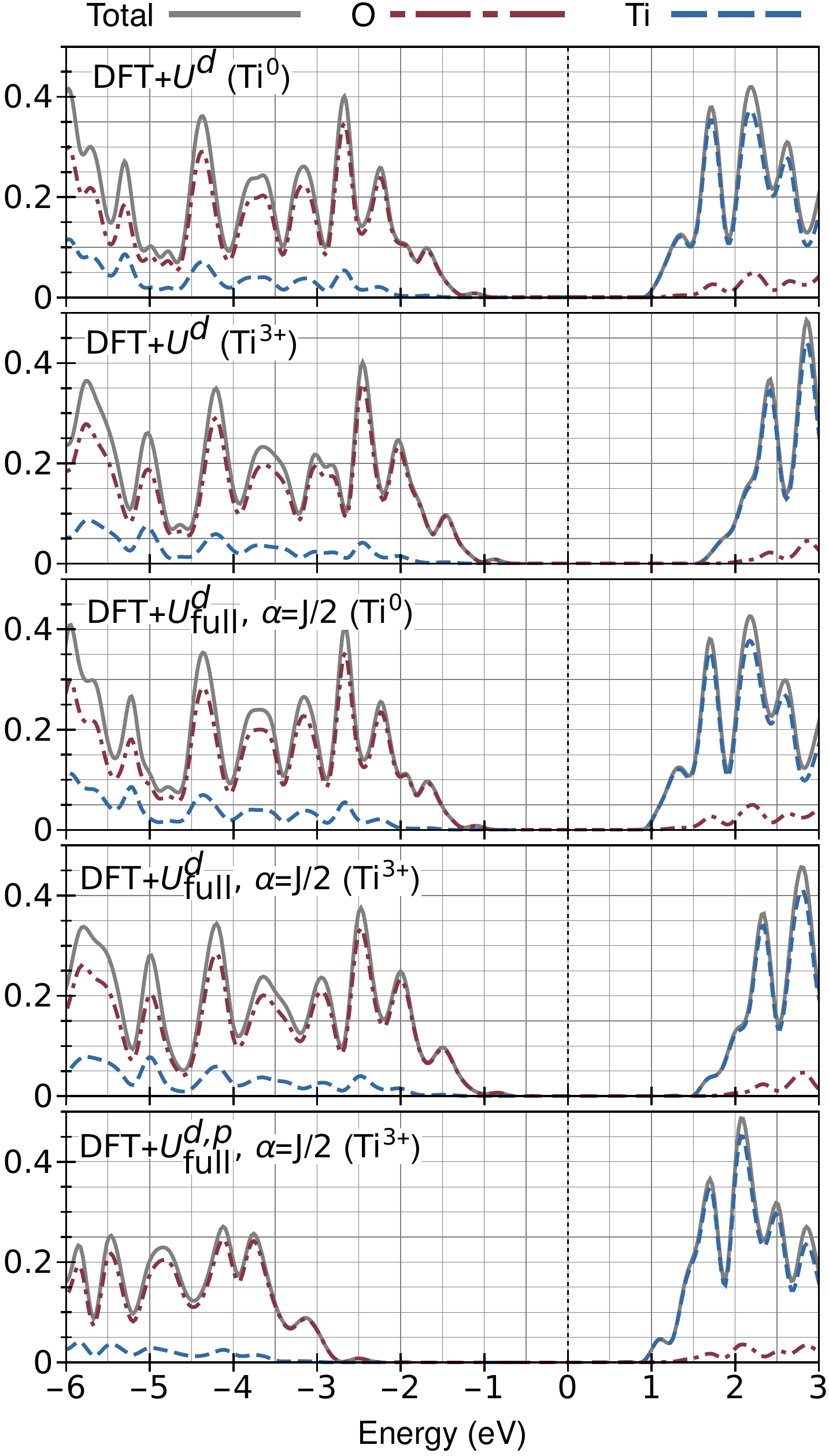}
\caption{The total and  local 
generalized Kohn-Sham density of states 
(LDOS) of pristine \rut 
calculated within DFT+$U^d$,
separately for Ti$^{0}$ and Ti$^{3+}$ subspace definitions,
within DFT+$U^d$+$J^d$ for the same two 
subspace definitions,
and finally within DFT+$U^{d,p}$+$J^{d,p}$ for
the Ti$^{3+}$ definition.
The spectrum is partitioned on a per-species basis using
Mulliken analysis based on the variationally optimized NGWFs.
Parameters where calculated from first principles
using the minimum-tracking linear-response method, 
and a Gaussian broadening of $0.1$~eV was used.
Each panel uses the mid-gap energy of the 
DFT+$U^d$ (Ti$^{0}$)
calculation for $0$~eV.}
\label{Fig2}
\end{figure}

\edit{For the specific case of rutile, we investigate in
detail here} the effect of varying the charge configuration for Ti used in the pseudo-atomic solver~\cite{PhysRevB.83.245124}, which constructs the set of the pseudo-atomic orbitals defining the $3d$ subspace of Ti.   
The neutral configuration is perhaps a natural choice, giving \edit{the relatively smooth,
diffuse subspace shown in 
Fig.~\ref{FigO}.
This results in less pressure on the plane-wave convergence} 
and, more importantly, 
it does not rely on any prior chemical intuition.
We also investigated the $3+$ 
atomic charge configuration, as a slightly more ``informed'' 
spatially localized subspace test case.
Given the LDA-appropriate $U$ and $J$ parameters 
calculated
for each of the two subspace types \edit{and presented
in Table~\ref{tab:rut-hub},} we performed the matching DFT+$U$, DFT+$U_\textrm{eff}$, and 
DFT+$U$+$J$ band-gap calculations, 
both  within \ud and \udup.
We also performed the ``cross'' calculations in the case of $\alpha = 0$,
i.e., where we used the $3+$ subspace parameters for
correcting the neutral subspace,
and vice-versa, in order to illustrate the separate effects
of over-localizating the projectors.

The results of these tests are shown in  Table~\ref{tab:rut-gap-Ti-conf-anal}.
We find that first-principles calculation of the Hubbard $U$ and Hund's $J$ parameters
does not compensate for the arbitrariness of the subspace choice, for  Ti $3d$. 
Instead, it reinforces this arbitrariness as far as the fundamental
gap is concerned in this system.
 Table~\ref{tab:rut-gap-Ti-conf-anal} reveals that this trend holds irrespective of whether
 correction is also applied to O $2p$ orbitals, denoted DFT+$U^{d,p}$, or
 indeed whether we are using DFT+$U$, DFT+$U_\textrm{eff}$, or DFT+$U_\textrm{full}$.
 As previously discussed, the increase in spatial localization of the $3d$ subspace, when
 we move from a neutral to a $3+$ configuration, increases the corresponding calculated $U$ and $J$
 parameters.
 This, of course, increases the predicted gap, when those parameters are applied to either 
 subspace type. %
 Moreover, \edit{Table~\ref{tab:rut-gap-Ti-conf-anal} demonstates}
  that, for either fixed set of parameters, the increase in
 subpspace localization also tends to open the gap, in this system, in fact by roughly the
 same amount.
 The net increase in the gap in going from \edit{the
 neutral to $3+$ subspace densities shown in Fig.~\ref{FigO}, 
 with corresponding first-principles parameters, is thus}
 approximately due, half-and-half, to the increase in parameters and increase in localization.
 
 On the basis of these results, we can envisage that   
 both the first-principles LDA-appropriate $U$ and $J$ parameters, and the fundamental gap
 for a fixed reasonable set of parameters, will attain maxima for some reasonable
 (though not generally the same) value of the pseudo-atomic configuration charge.
 A tentative step towards plotting observables as functions of a DFT+$U$
 subspace localization quantifier was presented in Ref.~\citenum{PhysRevB.82.081102}.
\edit{More recently, the projector 
dependence of DFT+$U$ results on rutile
TiO$_2$ has 
previously been demonstrated at fixed $U$ values in
Ref.~\citenum{doi:10.1021/acs.jctc.8b01211}.}
 We do not necessarily expect that the projector arbitrariness reinforcement effect
 will arise transition-metal oxides generally, 
particularly since projector arbitrariness cancellation has previously been observed in molecular FeO$^{+}$
using a self-consistently evaluated Hubbard $U$ parameter~\cite{doi:10.1063/1.2987444}.
This issue in DFT+$U$ clearly warrants further investigation on diverse systems using 
various approaches, such as parameter~\cite{doi:10.1063/1.3489110,glenn-phdthesis} or 
projector~\cite{PhysRevB.82.081102,korotin} self-consistency.
Pragmatically, we have found, in our minimum-tracking linear-response
calculations to date, that using the simplest, neutral pseudo-atomic configuration 
for constructing the DFT+$U$ projections works well 
relative to more localised charged configurations.
This is irrespective of the 
pseudopotential generator reference state, 
which is a somewhat different, technical matter
related the transferability in norm-conserving pseudopotentials.
  We note, in passing, that there is a small discrepancy in the gap from \edit{Ti$^{3+}$}-only subspace explicit DFT+$U$+$J$
and the \edit{corresponding unmodified-}DFT+$U$-code  equivalent form with $\alpha = J/2$, reflecting that calculations 
with \edit{excessively} localised subspaces are \edit{typically
less numerically stable,} aside from giving less favorable results.

{\clearpage


\section*{Supplementary material}

# Supplementary Material to
# "First-principles Hubbard $U$ and Hund's $J$ corrected approximate density-functional theory predicts an accurate fundamental gap in rutile and anatase $TiO_2$"

Okan K. Orhan[1] and David D. O'Regan[1]

[1]*School of Physics, AMBER, and CRANN Institute, Trinity College Dublin, the University of Dublin, Ireland*

In this Supplementary Material, we describe an investigation into how the first-principles DFT+$U$+$J$ correction, when applied to both Ti $3d$ and O $2p$ subspaces, affects the predicted internal ionic geometry and equilibrium unit cell volume of rutile and anatase $TiO_2$. We first, however, describe the Computational Details used overall, i.e., both here and in the main text.

## I. COMPUTATIONAL DETAILS

Initial crystallographic information for $TiO_2$-rutile and $TiO_2$-anatase were adopted from Refs. 1 and 2. Norm-conserving scalar-relativistic LDA pseudo-potentials were produced using the pseudo-potential generator OPIUM[3]. Following transferability testing, a $Ti^{3+}$ configuration with semi-core $3s$ and $3p$ orbitals in the valence and with relatively small, demanding cut-off radii of 1.54 $a_0$ (for $s$), 1.70 $a_0$ ($p$), and 1.82 $a_0$ ($d$), was chosen. Full geometry relaxation were performed with variable cell parameters at a high plane-wave cut-off energy $E_{cut}$ (75 Ha or approximately 2040 eV) and automatically generated $3 \times 3 \times 5$ ($TiO_2$-rutile) and $5 \times 5 \times 3$ (conventional-cell $TiO_2$-anatase) Monkhorst-Pack Brillouin zone sampling grids using the Quantum Espresso (QE) code[4,5]. The converged cell parameters were used to construct unfolded super-cells with 270 and 900 atoms, respectively, in order to emulate the same Brillouin zone sampling with real-valued Kohn-Sham orbitals within the ONETEP code[6]. This is a linear-scaling implementation of approximate KS-DFT using two nested optimization loops for the density kernel and a minimal set of non-orthogonal generalized Wannier functions (NGWFs)[7,8].

An under-pinning basis of $3^1\ 5^2 = 75$ *psinc* functions in all directions for $TiO_2$-rutile (corresponding to the effective kinetic energy cut-offs of $\sim 1776$ eV, $\sim 1776$ eV, and $\sim 1552$ eV), and $3^1\ 5^2 = 75$ in the $x$- and $y$-directions and $3^1\ 7^2 = 147$ in $z$-direction for $TiO_2$-anatase (corresponding to $\sim 945$ eV, $\sim 945$ eV, and $\sim 1578$ eV respectively), provided a basis set associated error of $\leq 1$ meV in the total energy per atom. A total of 13 variationally optimized NGWFs initially centred on Ti atoms were used, to complete the second and third periods up to Kr, and a total 4 NGWFs variationally optimized NGWFs initially centred on O atoms to complete the period up to Ar. A converged, common NGWF cut-off radius of 12 $a_0$ was used for both species, with the same total energy tolerance. For the band gaps reported in the main

text but not here in the Supplementary Material, a second set of NGWFs, with the number and cut-off radii, where then added and variationally optimised, following Ref. 9, in order to reproduce the Kohn-Sham states around the conduction-band minimum. For $Ti^0$ NGWF initial guesses, this makes a only small improvement to the predicted gaps, typically on the order of 5 meV, but the effect is significant (order of 0.1 eV) for $Ti^{3+}$ initial guesses. We emphasise that, for energies much above the conduction-band edges, the conduction band parts of the LDOS plots presented in the main text are qualitatively but not necessarily quantitatively reliable.

A discrete perturbation strength grid, $\delta\alpha^\uparrow = \{0, \pm0.01, \pm0.10, \pm0.50, \pm1.00\}$ eV was used to calculate the Hubbard $U$ and Hund's $J$ parameters, without restarts in order to remove any risk of premature convergence declaration and hence under-estimated response. This was applied to a single spin channel only, following the $2 \times 2$ procedure introduced in Ref. 10. A smooth response was obtained for all matrix elements for both species, Ti $3d$ and O $2p$, and for both crystal structures.

## II. EFFECTS OF DFT+$U$+$J$ ON IONIC GEOMETRY

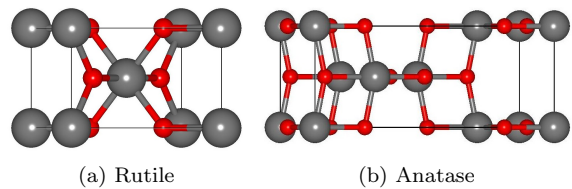

(a) Rutile          (b) Anatase

FIG. 1. Schematic crystal structures of $TiO_2$-rutile and $TiO_2$-anatase.

### A. Geometry optimization procedure

For the purposes only of this Supplemental Material, the same QE LDA geometry optimization procedure was repeated for $TiO_2$-rutile and $TiO_2$-anatase, using $10^{-7}$ Ry, $10^{-5}$ Ry/Bohr and $10^{-10}$ Ry convergence thresholds for the total energy, forces for ionic minimization, and self-consistency, respectively, but with fixed lattice vectors corresponding to $\pm1\%$ and $\pm2\%$ isotropic strain. As ONETEP is not equipped to perform cell-wall stress calculations, these optimized, strained unit cells

were then repeated to form supercells with the same atom counts as before for each polymorph, in order to carry out ONETEP internal ionic geometry optimizations using DFT (LDA) and DFT+$U$+$J$. In order to maintain a plane-wave equivalent real-space sampling of the supercell in ONETEP that remained commensurate with the unit-cell, however, it is necessary to conserve the number of real-space sampling (psinc) points. This means that, when a strain of $\pm x\%$ is applied, for $x \ll 1$, the equivalent plane-wave kinetic energy cut-off in each direction changes by a factor of $\approx \mp 2x\%$. While this effect is present to some extent in all plane-wave DFT codes, here it hampers more severely our ability to make precise conclusions as to the predicted unit cell volume.

Only the most successful DFT+$U$+$J$ functional in terms of the band gap was tested, in the expedient form DFT+$U_{full}^{d,p}$, $\alpha = J/2$. Only the first-principles parameters described in the main text were used, and we have not investigated the lattice volume dependence of the $U$ and $J$ parameters. Convergence thresholds of $5\times10^{-3}$ $a_0$, $10^{-6}$ Ha, and $10^{-4}$ Ha $a_0^{-1}$ were applied to the atomic displacement, total energy per atom, and maximum individual atomic force, respectively, within ONETEP. The energy versus volume data for each set of simulations were used to fit the Murnaghan equation-of-state[11], from which we extracted minimum-energy volumes and bulk moduli using a post-processing tool available within QE.

### B. Structural information for TiO$_2$-rutile

For rutile, Fig. 2 indicates a close match between the energy, and separately the band gap, changes with respect to isotropic strain across LDA (two codes) and DFT+$U$+$J$ (ONETEP, with fixed parameters). The total energies are not the same, differing by constants that are set to zero for illustrative purposes. The predicted band gap decreases with increasing lattice volume, albeit that this might feasibly not be the case if the parameters were made dependent upon the volume. Table I indicates that the small code dependence does not manifest in the internally relaxed bond lengths, and essentially that DFT+$U$+$J$ closely preserves the internal geometry of this direct-gap system at its LDA parameters.

Table II confirms that DFT+$U$+$J$ only marginally changes the predicted lattice volume and bulk modulus, subject to that caveat here that we cannot separately relax the three lattice vectors in ONETEP and we only explore isotropic strain with respect to the LDA (QE) relaxed unit cell. Remarkably, this result reflects those reported for very different, small molecular systems studied with a similar method in Ref. 10, where it was suggested that the O $2p$ correction compensates for Ti $3d$ correction in terms of the bonding strength and ionic forces. Finally, in order to compensate for discrepancy between the QE and ONETEP volumes, which most likely arises due to their differing plane wave grids (an approximate sphere of **G** vectors in the former case, a cuboid in the latter

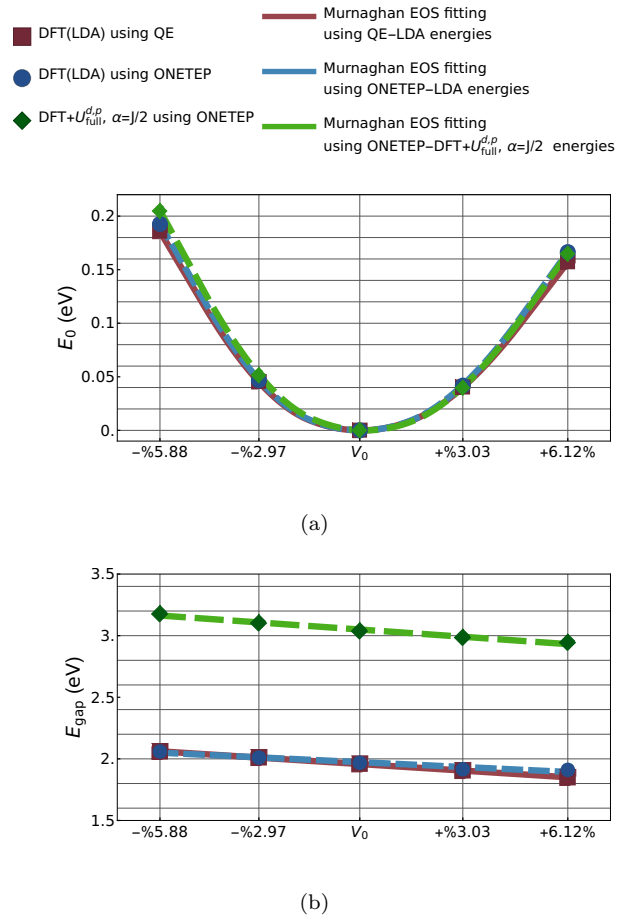

(a)

(b)

FIG. 2. (a) Total energies (per formula unit cell) as a function of isotropically-strained (QE LDA) lattice volume of TiO$_2$-rutile calculated using LDA within Quantum Espresso and ONETEP, and also using first-principles DFT+$U$+$J$ within ONETEP, each with their respective converged internal ionic geometries. (b) The Kohn-Sham, or generalized Kohn-Sham, band gaps corresponding to (a) are shown here with respective linear fits. In the case of ONETEP data, there is an uncontrolled overestimation $\sim 1$ meV introduced here with respect to the results in the main text, due to non-optimization of conduction-band NGWFs.

case), we define the 'code corrected' as the Murnaghan interpolated DFT+$U$+$J$ volume, multiplied by the ratio of the interpolated LDA (QE) and LDA (ONETEP) volumes. Extrapolating the band gap to this volume, and also applying the DFT+$U$+$J$ conduction-band optimization correction calculated at the LDA (QE) geometry, we arrive at the DFT+$U$+$J$ (corrected) band gap. For rutile, this is the same as the uncorrected result to the reasonable number of significant figures shown, which is also the same as the LDA (QE) geometry DFT+$U$+$J$ shown in the main text. For rutile TiO$_2$, therefore, perhaps as a consequence of its direct band gap or relatively similar lattice vector lengths, the consequences for the geometry of changing the plane-wave code used, or changing from the LDA to first-principles DFT+$U$+$J$ (with O $2p$

| | Lattice vectors ($a_0$) | | |
|---|---|---|---|
| $\mathbf{v}_1^0$ | 8.5289 | 0.0000 | 0.0000 |
| $\mathbf{v}_2^0$ | 0.0000 | 8.5289 | 0.0000 |
| $\mathbf{v}_3^0$ | 0.0000 | 0.0000 | 5.4729 |

| | Atomic positions (fractional) | | |
|---|---|---|---|
| Ti | 0.0000 | 0.0000 | 0.0000 |
| Ti | 0.5000 | 0.5000 | 0.5000 |
| O | 0.3040 | 0.3040 | 0.0000 |
| O | 0.6960 | 0.6960 | 0.0000 |
| O | 0.1960 | 0.8041 | 0.5000 |
| O | 0.8041 | 0.1960 | 0.5000 |

| | Bond lengths ($a_0$) | |
|---|---|---|
| | Ti – O (4 of) | Ti – O (2 of) |
| LDA (QE) | 3.6160 | 3.6671 |
| LDA (ONETEP) | 3.6160 | 3.6671 |
| DFT+$U$+$J$ | 3.6141 | 3.6700 |
| Experiment[12] | 3.6772 | 3.7349 |

TABLE I. Optimized zero-strain internal ionic geometries of $TiO_2$-rutile, as predicted by two implementations of DFT (LDA) and by first-principles DFT+$U$+$J$ in the guise of DFT+$U_{\mathrm{full}}^{d,p}, \alpha = J/2$, alongside experimental values for comparison. The QE LDA lattice vectors are used in all cases.

correction) are numerically small.

## C. Structural information for $TiO_2$-anatase

The procedure and data representation of the previous sub-section is repeated here for the case of anatase. In Table III, we first note that one of the lattice vectors is over twice the length of the others, which gives rise to a more significantly non-uniform equivalent plane-wave grid in ONETEP. Nonetheless, as the potential is well converged, we note only a small difference in internally-relaxed LDA bond lengths of QE and ONETEP. The change to DFT+$U$+$J$ makes a larger but still small change, and does not significantly improve the bond lengths with respect to their experimental values.

Turning next to, Fig. 3, we observe that for anatase the effect of the DFT+$U$+$J$ correction on the predicted equilibrium lattice volume (subject always to the caveat that only isotropic strain is considered) is rather small compared to the effect of changing the code. We expect that this arises due to the aforementioned more significant dependence of the effective plane-wave cut-off energy on the lattice volume in ONETEP, when compared to QE, which is made more apparent due to the very anisotropic lattice vectors of anatase. Again, as the potential is well

| | $V_0^{\mathrm{fit}}$ ($a_0^3$) | % Error | $K$ (GPa) |
|---|---|---|---|
| LDA (QE) | 398.26 | -5.50 | 256.8 |
| LDA (ONETEP) | 398.27 | -5.50 | 269.9 |
| DFT+$U$+$J$ | 398.69 | -5.40 | 276.8 |
| DFT+$U$+$J$ (corrected) | 398.68 | -5.40 | - |
| Experiment[13] | 421.43[a] | - | 211.5[a] |

| | Evaluated at $V_0$ | |
|---|---|---|
| | $E_{\mathrm{gap}}$ (eV) | $dE_{\mathrm{gap}}^{\mathrm{fit}}/dV$ (eV $a_0^{-3}$) |
| LDA (QE) | 1.96 | $-4.4 \times 10^{-3}$ |
| LDA (ONETEP) | 1.97 | $-3.3 \times 10^{-3}$ |
| DFT+$U$+$J$ | 3.04 | $-4.9 \times 10^{-3}$ |
| DFT+$U$+$J$ (corrected) | 3.04 | - |

TABLE II. Estimated minima $V_0^{\mathrm{fit}}$ of Murnaghan equation of state, as fitted to the ground-state energies of $TiO_2$-rutile using LDA within QE, and using LDA and DFT+$U$+$J$ within ONETEP (top panel), and in all cases based on strained LDA (QE) lattice vectors but with internal ionic coordinates relaxed with the stated functional. The percentage volume error with respect to experiment is provided, as well as an estimate of the bulk modulus ($K$). Also shown are the calculated band gaps for the internally-relaxed geometry at the LDA (QE) volume $V_0$, as well as the slope of the band gap with respect to volume as extracted from a linear fit. DFT+$U$+$J$ (corrected) values are our best estimate, as extrapolated using the linear fit based on the 'code corrected' DFT+$U$+$J$ volume described in the text, and applying the known correction constant for conduction-band edge optimization calculated at the LDA (QE) geometry.

converged, the effect on the LDA band gap of changing the code is small. The predicted gaps decrease with lattice volume, but at a lower rate than that found which we observed in rutile.

Turning finally to Table IV, we note that DFT+$U$+$J$ causes an expansion of the volume in anatase, when the code is approximately corrected for, and although this is, as for rutile, a change in the direction of the experimental value, the change is very small. When internal geometric relaxation is included, with the LDA (QE) lattice vectors, the predicted DFT+$U$+$J$ band gap increases by 0.06 eV with respect to the value reported in the main text, or by 0.04 eV when anisotropic plane-wave basis and conduction-band optimization related errors are approximately corrected. This takes us further from the most reliable available experimental fundamental gap value of 3.47 eV, and, noting that the bond lengths change only marginally, it is not clear whether this due to a genuine geometry relaxation effect on the gap. Alternative possibilities include a gap-preserving shift in indirect band edge wave-vectors, which would result in the approximately sampled gap spuriously opening, or the

| | Lattice vectors ($a_0$) | | |
|---|---|---|---|
| $\mathbf{v}_1^0$ | 7.0131 | 0.0000 | 0.0000 |
| $\mathbf{v}_2^0$ | 0.0000 | 7.0131 | 0.0000 |
| $\mathbf{v}_3^0$ | 0.0000 | 0.0000 | 17.7311 |

| | Atomic positions (fractional) | | |
|---|---|---|---|
| Ti | 0.0000 | 0.2500 | 0.3750 |
| Ti | 0.0000 | 0.7500 | 0.6250 |
| Ti | 0.5000 | 0.7500 | 0.8750 |
| Ti | 0.5000 | 0.2500 | 0.1250 |
| O | 0.0000 | 0.2500 | 0.1670 |
| O | 0.0000 | 0.7500 | 0.8330 |
| O | 0.5000 | 0.7500 | 0.6670 |
| O | 0.5000 | 0.2500 | 0.3330 |
| O | 0.0000 | 0.7500 | 0.4170 |
| O | 0.0000 | 0.2500 | 0.5830 |
| O | 0.5000 | 0.7500 | 0.0830 |
| O | 0.5000 | 0.2500 | 0.9170 |

| | Bond lengths ($a_0$) | |
|---|---|---|
| | Ti – O (4 of) | Ti – O (2 of) |
| LDA (QE) | 3.5851 | 3.6865 |
| LDA (ONETEP) | 3.5851 | 3.6855 |
| DFT+$U$+$J$ | 3.5898 | 3.6642 |
| Experiment[12] | 3.6513 | 3.7394 |

TABLE III. Optimized zero-strain internal ionic geometries of $TiO_2$-anatase, as predicted by two implementations of DFT (LDA) and by first-principles DFT+$U$+$J$ in the guise of DFT+$U_{\text{full}}^{d,p}$, $\alpha = J/2$, alongside experimental values for comparison. The QE LDA lattice vectors are used in all cases.

real-space grid points shifting to locations where some of the DFT+$U$+$J$ projectors have a greater amplitude, again giving rise to an increased gap. The former effect would be identified by non-self-consistent band-structure interpolation, were it technically available, and the latter effect, if present, might be mitigated by geometry self-consistent calculated $U$ and $J$ parameters.

Overall, subject to the technical limitations of our study, we can tentatively conclude that in the systems investigated, first-principles DFT+$U$+$J$ with O $2p$ correction makes only a marginal change to the ionic geometries of the underlying LDA, and that the effects of changing the DFT code used can be more significant. As we have developed a prescription for exactly emulating DFT+$U$+$J$ for closed-shell systems in codes with no Hund's $J$ implementation (using the parameters $U_{\text{full}} = U - 2J$ and $\alpha = J/2$), the stage is now set for future, lattice and internal geometry self-consistent DFT+$U$+$J$ calculations with finely sampled band-structures.

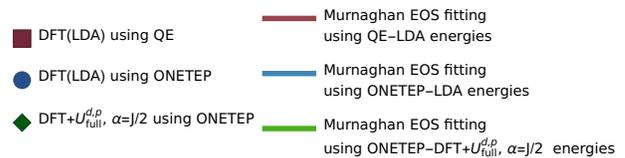

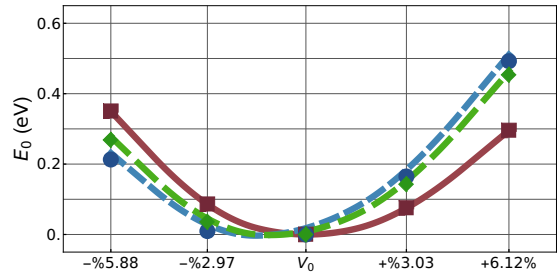

(a)

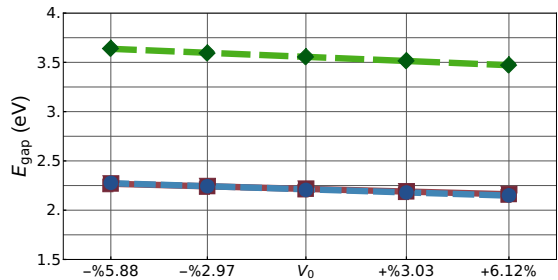

(b)

FIG. 3. (a) Total energies (per formula unit cell) as a function of isotropically-strained (QE LDA) lattice volume of $TiO_2$-anatase calculated using LDA within Quantum Espresso and ONETEP, and also using first-principles DFT+$U$+$J$ within ONETEP, each with their respective converged internal ionic geometries. (b) The Kohn-Sham, or generalized Kohn-Sham, band gaps corresponding to (a) are shown here with respective linear fits. In the case of ONETEP data, there is an uncontrolled overestimation $\sim 1$ meV introduced here with respect to the results in the main text, due to non-optimization of conduction-band NGWFs.

|  | $V_0^{\text{fit}}$ ($a_0^3$) | % Error | $K$ (GPa) |
|---|---|---|---|
| LDA (QE) | 872.52 | -4.89 | 220.9 |
| LDA (ONETEP) | 860.45 | -6.20 | 253.2 |
| DFT+$U$+$J$ | 864.09 | -5.81 | 255.3 |
| DFT+$U$+$J$ (corrected) | 876.21 | -4.49 |  |
| Experiment | 917.37[14] | - | 178[15] |

| Evaluated at $V_0$ | $E_{\text{gap}}$ (eV) | $dE_{\text{gap}}^{\text{fit}}/dV$ (eV $a_0^{-3}$) |
|---|---|---|
| LDA (QE) | 2.22 | - 1.0 $\times 10^{-3}$ |
| LDA (ONETEP) | 2.21 | - 1.2 $\times 10^{-3}$ |
| DFT+$U$+$J$ | 3.56 | - 1.6 $\times 10^{-3}$ |
| DFT+$U$+$J$ (corrected) | 3.54 | - |

TABLE IV. Estimated minima $V_0^{\text{fit}}$ of Murnaghan equation of state, as fitted to the ground-state energies of $TiO_2$-anatase using LDA within QE, and using LDA and DFT+$U$+$J$ within ONETEP (top panel), and in all cases based on strained LDA (QE) lattice vectors but with internal ionic coordinates relaxed with the stated functional. The percentage volume error with respect to experiment is provided, as well as an estimate of the bulk modulus ($K$). Also shown are the calculated band gaps for the internally-relaxed geometry at the LDA (QE) volume $V_0$, as well as the slope of the band gap with respect to volume as extracted from a linear fit. DFT+$U$+$J$ (corrected) values are our best estimate, as extrapolated using the linear fit based on the 'code corrected' DFT+$U$+$J$ volume described in the text, and applying the known correction constant for conduction-band edge optimization calculated at the LDA (QE) geometry.


}

\end{document}